%% file: main.tex
\documentclass[letterpaper]{article}

\usepackage{tikz}

\usepackage{natbib,alifeconf}  

\usepackage{hyperref}

\usepackage{subcaption}

\usepackage{hhline}

\usepackage{filecontents}
\usepackage{csvsimple}

\usepackage{amsmath}

\usepackage{rotating}

\usepackage[export]{adjustbox}

\ifdefined\mydraft
\mydraft
\fi

\graphicspath{{img/}}

\title{Exploring Evolved Multicellular Life Histories in a Open-Ended Digital Evolution System}
\author{Matthew Andres Moreno$^{1}$ \and Charles Ofria$^{1}$ \\
\mbox{}\\
$^1$BEACON Center, Michigan State University, East Lansing, MI 48824 \\
mmore500@msu.edu} 

\begin{document}
\maketitle

\input{tex/abstract.tex}

\input{tex/introduction.tex}

\input{tex/methods.tex}

\input{tex/results.tex}

\input{tex/conclusion.tex}

\input{tex/acknowledgements.tex}

\footnotesize
\bibliographystyle{apalike}
\bibliography{bibl} 

\clearpage
\newpage

\appendix

\input{tex/supplementary.tex}

\end{document}

%% file: tex/abstract.tex
\begin{abstract}

Evolutionary transitions occur when previously-independent replicating entities unite to form more complex individuals.
Such transitions have profoundly shaped natural evolutionary history and occur in two forms: fraternal transitions involve lower-level entities that are kin (e.g., transitions to multicellularity or to eusocial colonies), while egalitarian transitions involve unrelated individuals (e.g., the origins of mitochondria).
The necessary conditions and evolutionary mechanisms for these transitions to arise continue to be fruitful targets of scientific interest.
Here, we examine a range of fraternal transitions in populations of open-ended self-replicating computer programs.
These digital cells were allowed to form and replicate kin groups by selectively adjoining or expelling daughter cells.
The capability to recognize kin-group membership enabled preferential communication and cooperation between cells.
We repeatedly observed group-level traits that are characteristic of a fraternal transition.
These included reproductive division of labor, resource sharing within kin groups, resource investment in offspring groups, asymmetrical behaviors mediated by messaging, morphological patterning, and adaptive apoptosis.
We report eight case studies from replicates where transitions occurred and explore the diverse range of adaptive evolved multicellular strategies.
\end{abstract}

%% file: tex/introduction.tex
\section{Introduction}

An evolutionary transition in individuality is an event where independently replicating entities unite to replicate as a single, higher-level individual \citep{smith1997major}.
These transitions are understood as essential to natural history's remarkable record of complexification and diversification \citep{smith1997major}.
Likewise, researchers studying open-ended evolution in artificial life systems have focused on transitions in individuality as a mechanism that is missing in digital systems, but necessary for achieving the evolution of complexity and diversity that we witness in nature \citep{taylor2016open, banzhaf2016defining}.

Here, we focus on \textit{fraternal} evolutionary transitions in individuality, in which the higher-level replicating entity is derived from the combination of cooperating kin that have entwined their long-term fates \citep{west2015major}.
Multicellular organisms and eusocial insect colonies exemplify this phenomenon \citep{smith1997major} given that both are sustained and propagated through the cooperation of lower-level kin.
Although not our focus here, egalitarian transitions --- events in which non-kin unite, such as the genesis of mitochondria by symbiosis of free-living prokaryotes and eukaryotes \citep{smith1997major} --- also constitute essential episodes in natural history.

In nature, major transitions occur rarely and over vast time scales, making them challenging to study.
Recent work in experimental evolution \citep{ratcliff2014experimental, ratcliff2015origins, gulli2019evolution, koschwanez2013improved}, mechanistic modeling \citep{hanschen2015evolutionary, staps2019emergence}, and artificial life \citep{goldsby2012task, goldsby2014evolutionary} complements traditional post hoc approaches focused on characterizing the record of natural history.
These systems each instantiate the evolutionary transition process, allowing targeted manipulations to test hypotheses about the requisites, mechanisms, and evolutionary consequences of fraternal transitions.

Our work here follows closely in the intellectual vein of Goldsby's deme-based artificial life experiments \citep{goldsby2012task, goldsby2014evolutionary}.
In her studies, high-level organisms exist as a group of cells within a segregated, fixed-size subspace.
High-level organisms that must compete for a limited number of subspace slots.
Individual cells that comprise an organism are controlled by heritable computer programs that allow them to self-replicate, interact with their environment, and communicate with neighboring cells.

Goldsby's work defines two modes of cellular reproduction: tissue accretion and offspring generation.
Within a group, cells undergo tissue accretion, whereby a cell copies itself into a neighboring position in its subspace.
In the latter, a population slot is cleared to make space for a daughter organism then seeded with a single daughter cell from the parent organism.

Cells grow freely within an organism, but fecundity depends on the collective profile of computational tasks (usually mathematical functions) performed within the organism.
When an organism accumulates sufficient resource, a randomly chosen subspace is cleared and a single cell from the replicating organism is used as a propagule to seed the new organism.
This setup mirrors the dynamics of biological multicellularity, in which cell proliferation may either grow an existing multicellular body or found a new multicellular organism.

Here, we take several steps to develop a computational environment that removes the enforcement and rigid regulation of multiple organismal levels.
Specifically, we remove the explicitly segregated subspaces and we let cell interact with each other more freely.
We demonstrate the emergence of multicellularity where each organism manages its own spatial distribution and reproductive process.
This spatially-unified approach enables more nuanced interactions among organisms, albeit at the cost of substantially more complicated analyses.
Instead of a single explicit interface to mediate interactions among high-level organisms, such interactions must emerge via many cell-cell interfaces.
Novelty can occur in terms of interactions among competitors, among organism-level kin, or even within the building blocks that make up hierarchical individuality.
Experimentally studying fraternal transitions in a digital system where key processes (reproductive, developmental, homeostatic, and social) occur implicitly within a unified framework can provide unique insights into nature.

We do provide some framework to facilitate fraternal transitions in individuality by allowing cells to readily designate distinct kin groups.
Offspring cells may either remain part their parent's hereditary group or found a new group.
Cells can recognize group members, thus allowing targeted communication and resource sharing with kin.
We reward cells for performing tasks that require proper timing such that they must coordinate to be successful.
As such, cells that cooperate will have an advantage on those tasks and if they are also part of a hereditary group they will increase their inclusive fitness.
In previous work we evolved parameters for manually-designed cell-level strategies to explore fraternal transitions in individuality \citep{moreno2019toward}.
Here, we extend this prior work to use a more open-ended event-driven genetic programming representation called SignalGP, which was designed to facilitate dynamic interactions among agents and between agents and their environment \citep{lalejini2018evolving}.  As expected, we see a far more diverse set of behaviors and strategies arise.

%% file: tex/methods.tex
\section{Methods}

We performed simulations in which cells evolved open-ended behaviors to make decisions about resource sharing, reproductive timing, and apoptosis.
We will first describe the environment and hereditary grouping system cells evolved under and then describe the behavior-control system cells used.

\subsection{Environment and Hereditary Groups}

Cells occupy individual tiles on a toroidal grid.
Over discrete time steps (``updates''), cells can collect a resource.
Once sufficient resource has been accrued, cells may pay one unit of resource to place a daughter cell on an adjoining tile of the toroidal grid (i.e., reproduce), replacing any existing cell already there.
Collected resource decays at a rate of 0.1\% per update, incentivizing its quick use.

Cells accrue resource via a cooperative resource-collection process conducted by explicitly-registered hereditary groups.
As cells reproduce, they can choose to keep offspring in the parent's hereditary group or expel offspring to found a new hereditary group.
These decisions affect the spatial layout of these hereditary groups and, in turn, affect individual cells' resource-collection rate.
Medium-sized, circular hereditary groups tend to collect resource at a greater per-cell rate than large, small, or irregularly-shaped groups.
To promote group turnover, we counteract the advantage of established hereditary groups with a simple aging scheme.
As hereditary groups age over elapsed updates and somatic generations, their constituent cells expressly lose the ability regenerate somatic tissue and then, soon after, to collect resource.

A complete description of the mechanisms behind these collective resource-collection and group aging mechanisms appears in supplementary sections \ref{sup:resource_collection_process} and \ref{sup:channel_group_life_cycle}.

Because hereditary groups arise through inheritance, they signify a kin relationship in addition to a potentially functionally cooperative relationship.
In this work, we screen for fraternal transitions in individuality with respect to these hereditary groups by evaluating three characteristic traits of higher-level organisms: resource sharing, reproductive division of labor, and apoptosis.

\subsection{Cell-Level Organisms}

Our experiments use cell-level digital organisms controlled by genetic programs subject to mutations and implicit selective pressures.
We employ the SignalGP representation, which expresses function-like modules of code in response to internal signals or external stimuli (akin to gene expression).
This event-driven paradigm facilitates the evolution of dynamic interactions between digital organisms and their environment (including other organisms) \citep{lalejini2018evolving}.

Previous work evolving digital organisms in grid-based problem domains has relied on a single computational agent that designates a direction to act in via an explicit cardinal ``facing'' \citep{goldsby2014evolutionary, goldsby2018serendipitous, grabowski2010early, biswas2014causes, lalejini2018evolving}.
We introduce novel methodology to facilitate the evolution of directionally-symmetric phenotypes.
In this work, each cell instantiates four copies of the SignalGP hardware: one facing each cardinal direction.
These hardware instances all execute the same SignalGP program and may coordinate via internal signals, but are otherwise decoupled.
Supplementary Figure \ref{fig:dishtinygp-cartoon} overviews the configuration of the four SignalGP instances that constitute a single cell.

\subsection{Surveyed Evolutionary Conditions}

To broaden our exploration of possible evolved multicellular behaviors in this system, we surveyed several evolutionary conditions.

In one manipulation, we explored the effect of structuring hereditary groups, such that parent cells can choose to keep offspring in their same sub-group, in just the same full group, or expel them entirely to start a new group.
Cells can independently mediate their behavior based on the level of the group with which they are interacting.

In a second manipulation, we explored the importance of explicitly selecting for medium-sized groups (as had been needed to maximize resource collection) by removing this incentive.
Instead, they system distributed resource at a uniform per-cell rate.

We combined these two manipulations to yield four surveyed conditions:
\begin{enumerate}
\item ``Flat-Even'': One hereditary level (flat) with uniform resource inflow (even). In-browser simulation: \url{https://mmore500.com/hopto/i},
\item ``Flat-Wave'': One hereditary level (flat) with group-mediated resource collection (wave); In-browser simulation: \url{https://mmore500.com/hopto/j}),
\item ``Nested-Even'': Two hierarchically-nested hereditary levels (nested) with uniform resource inflow (even). In-browser simulation: \url{https://mmore500.com/hopto/k},
\item ``Nested-Wave'': Two hierarchically-nested hereditary group levels (nested) with group-mediated resource collection (wave). In-browser simulation: \url{https://mmore500.com/hopto/l}.
\end{enumerate}

%% file: tex/results.tex
\section{Results and Discussion}

To characterize the general selective pressures induced by surveyed environmental conditions, we assessed the prevalence of characteristic multicellular traits among evolved genotypes across replicates.
In the case of an evolutionary transition of individuality, we would expect cells to modulate their own reproductive behavior to prioritize group interests above individual cell interests.
In DISHTINY, cell reproduction inherently destroys an immediate neighbor cell.
As such, we would expect somatic growth to occur primarily at group peripheries in a higher-level individual.
Supplementary Figure \ref{fig:reproduction_surrounded} compares cellular reproduction rates between the interior and exterior of highest-level same-channel signaling networks.
For all treatments, phenotypes with depressed interior cellular reproduction rates dominated across replicates (non-overlapping 95\% CI).
All four treatment conditions appear to select for some level of reproductive cooperation among cells.

Across replicate evolutionary runs in all four treatments, we also found that resource was transferred among registered kin at a significantly higher mean rate than to unrelated neighbors (non-overlapping 95\% CI).
To test whether this resource-sharing was solely an artifact of sharing between direct cellular kin, we also assessed mean sharing to registered kin that were not immediate cellular relatives.
Mean sharing between such cells also exceeded sharing among unrelated neighbors (non-overlapping 95\% CI).
Thus, all four treatments appear to select for functional cooperation among wider kin groups.
Supplementary section \label{sec:resource-sharing} presents these results in detail.

\subsection{Qualitative Life Histories} \label{sec:life-histories}

\input{fig/lifecycle.tex}

Although cooperative cell-level phenotypes were common among evolved same-channel signaling networks, across replicates functional and reproductive cooperation arose via diverse qualitative life histories.
To provide a general sense for the types of life histories we observed in this system, Figure \ref{fig:lifecycle} shows time lapses of representative multicellular groups evolved in different replicates.
Figure \ref{fig:lifecycle-naive} depicts an example of a naive life history in which --- beyond the cellular progenitor of a propagule group --- the parent and propagule groups exhibit no special cooperative relationship.
In Figure \ref{fig:lifecycle-adjoin}, propagules repeatedly bud off of parent groups to yield a larger network of persistent parent-child cooperators.
In Figure \ref{fig:lifecycle-sweep}, propagules are generated at the extremities of parent groups and then rapidly replace most or all of the parent group.
Finally, in Figure \ref{fig:lifecycle-burst}, propagules are generated at the interior of a parent group and replace it from the inside out.

To better understand the multicellular strategies that evolved in this system, we investigated the mechanisms and adaptiveness of notable phenotypes that evolved in several individual evolutionary replicates.
In the following sections, we present these investigations as a series of case studies.


\subsection{Case Study: Burst Lifecycle} \label{sec:gene-regulation}

\input{fig/ko-interior_propagule.tex}


In the strain exhibiting the ``burst'' lifecycle shown above, we wondered how localization and timing of the propagule origination was determined.
To assess whether gene regulation instructions played a role in this process, we prepared two knockout strains.
In the first, gene regulation instructions were replaced with no-operation (Nop) instructions (so that gene regulation state would remain baseline).
In the second, the reproduction instructions to spawn a propagule were replaced with Nop instructions.
Figure \ref{fig:regulation_visualizations} depicts the gene regulation phenotypes of these strains.

Figure \ref{fig:interior_propagule_rate} compares interior propagule generation between the strains, confirming the direct mechanistic role of gene regulation in promoting interior propagule generation (non-overlapping 95\% CI).

In head-to-head match-ups, the wild type strain outcompetes both the regulation-knockout ($20/20$; $p < 0.001$; two-tailed exact test) and the propagule-knockout strains
($20/20$; $p < 0.001$; two-tailed exact test).
The deficiency of the propagule-knockout strain confirms the adaptive role of interior propagule generation.
Likewise, the deficiency of the regulation-knockout strain affirms the adaptive role of gene regulation in the focal wild type strain.

\subsection{Case Study: Cell-cell Messaging} \label{sec:cell-cell-messaging}

\input{fig/ko-intermessaging-sharing.tex}

We discovered adaptive cell-cell messaging in two evolved strains.
Here, we discuss a strain evolved under the Flat-Wave treatment where cell-cell messaging disrupts directional and spatial uniformity of resource sharing.
Supplementary Section \ref{sec:intergroup} overviews an evolved strain where cell-cell messaging appears to intensify expression of a contextual tit-for-tat policy between same-channel signaling network groups.

Figure \ref{fig:ko-intermessaging-sharing} depicts the cell-cell messaging, resource sharing, and resource stockpile phenotypes of the wild type strain side-by-side with corresponding phenotypes of a cell-cell messaging knockout strain.
In the wild type strain, cell-cell messaging emanates from irregular collection of cells --- in some regions, grid-like and in others more sparse --- broadcasting to all neighboring cells.
Resource sharing appears more widespread in the knockout strain than in the wild type.
However, messaging's effects suppressing resource sharing is neither spatially nor directionally homogeneous.
Relative to the knockout strain, cell-cell messaging increases variance in cardinal directionality of net resource sharing
(%
WT: mean 0.28, S.D. 0.07, $n=54$; 
KO: mean 0.17, S.D. 0.07, $n=69$; 
$p < 0.001$, bootstrap test%
).
Cell-cell messaging also increases variance of resource sharing density with respect to spatial quadrants demarcated by same-channel signaling group's spatial centroid
(%
WT: mean 0.23, S.D. 0.07, $n=52$; 
KO: mean 0.16, S.D. 0.08, $n=68$; 
$p < 0.001$, bootstrap test
).
We used competition experiments to confirm the fitness advantage both of cell-cell messaging ($20/20$; $p < 0.001$; two-tailed exact test) and (using a separate knockout strain) resource sharing ($20/20$; $p < 0.001$; two-tailed exact test).
The fitness advantage of irregularized sharing might stem from a corresponding increase in the fraction of cells with enough resource to reproduce stockpiled
(%
WT: mean 0.18, S.D. 0.11, $n=54$; 
KO: mean 0.06, S.D. 0.08, $n=69$; 
$p < 0.001$, bootstrap test
).

\subsection{Case Study: Gradient-conditioned Cell Behavior} \label{sec:gradient-conditioned-behavior}

\input{fig/ko-stockpiletrigger-sharing.tex}

To further assess how multicellular groups process and employ spatial and directional information, we investigated whether successful multicellular strategies evolved where cells condition their behavior based on the resource concentration gradient within a multicellular group.
We discovered a strain that employs a dynamic strategy where cells condition their own resource-sharing behavior based on the relative abundance of their own resource stockpiles compared to their neighbors.
This strain appears to use this information to selectively suppress resource sharing.
This strain's wild type outcompeted a variant where cells' capacity to assess relative richness of neighboring resource stockpiles was knocked out ($20/20$; $p < 0.001$; two-tailed exact test).
Figure \ref{fig:ko-stockpiletrigger-sharing} contrasts the wild type resource-sharing phenotype with the more sparse knockout resource-sharing phenotype.

This result raises the question of whether more sophisticated morphological patterning might evolve within the experimental system.
Supplementary Section \ref{sec:morphology} examines a strain that exhibited genetically-driven morphological patterning of same-channel signaling networks.
However, in knockout experiments this phenotype did not exhibit adaptive significance.

\subsection{Case Studies: Apoptosis} \label{sec:apoptosis}

\input{fig/ko-apoptosis.tex}

Finally, we assessed whether cell self-sacrifice played a role in multicellular strategies evolved across our survey.
Screening replicate evolutionary runs by apoptosis rate flagged two strains with several orders of magnitude greater activity.
In strain A, evolved under the Nested-Even treatment, apoptosis accounts for 2\% of cell mortality.
In strain B, evolved under the Nested-Flat treatment, 15\% of mortality is due to apoptosis.

To test the adaptive role of apoptosis in these strains, we performed competition experiments against apoptosis knockout strains, in which all apoptosis instructions were substituted for Nop instructions.
Figure \ref{fig:ko-apoptosis} compares the same-channel wild type phenotypes of these strains to their corresponding knockouts.

Apoptosis contributed significantly to fitness in both strains (strain A: $18/20$, $p < 0.001$, two-tailed exact test; strain B: $20/20$, $p < 0.001$, two-tailed exact test).
The success of strategies incorporating cell suicide is characteristic of evolutionary conditions favoring altruism, such kin selection or a transition from cell-level to collective individuality.

To discern whether spatial or temporal targeting of apoptosis contributed to fitness, we competed wild type strains with apoptosis-knockout strains on which we externally triggered cell apoptosis with spatially and temporally uniform probability.
In one set of competition experiments, the knockout strain's apoptosis probability was based on the observed apoptosis rate of the wild type strain's monoculture.
In a second set of competition experiments, the knockout strain's apoptosis probability was based on the observed apoptosis rate of the population in the evolutionary run the wild type strain was harvested from.
In both sets of experiments on both strains, wild type strains outcompeted knockout strains with uniform apoptosis probabilities
(%
strain A \@ monoculture rate: $18/20$, $p < 0.001$, two-tailed exact test;
strain A \@ population rate: $19/20$, $p < 0.001$, two-tailed exact test;
strain B \@ monoculture rate: $20/20$, $p < 0.001$, two-tailed exact test;
strain B \@ population rate: $20/20$, $p < 0.001$, two-tailed exact test%
). 

%% file: fig/lifecycle.tex
\begin{figure*}[!htbp]
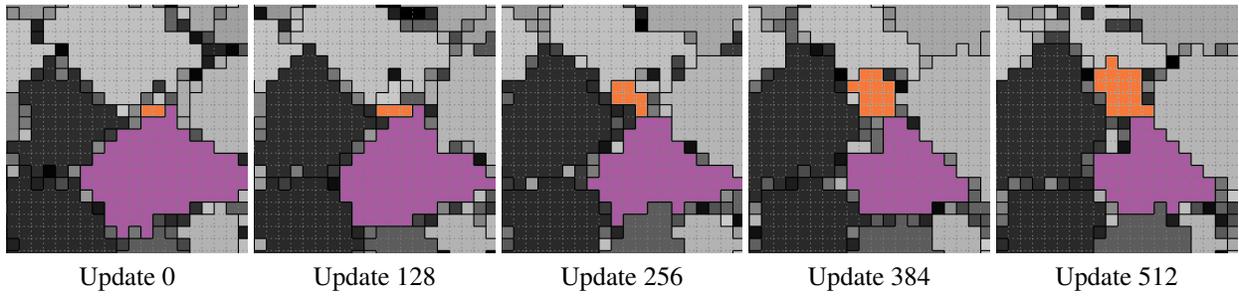
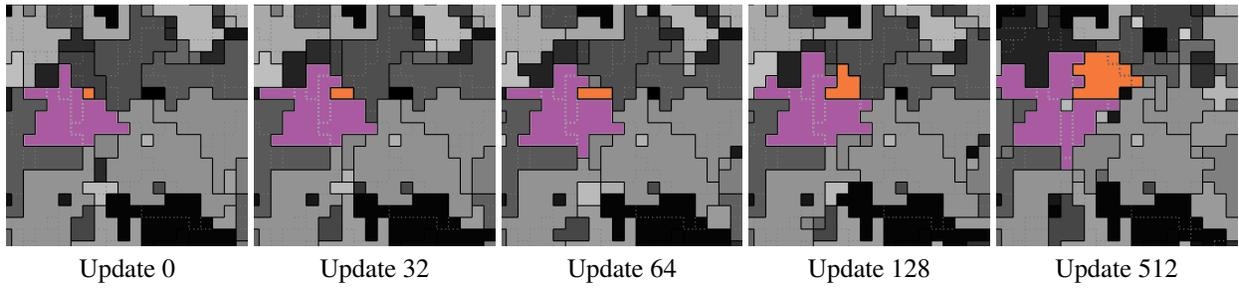
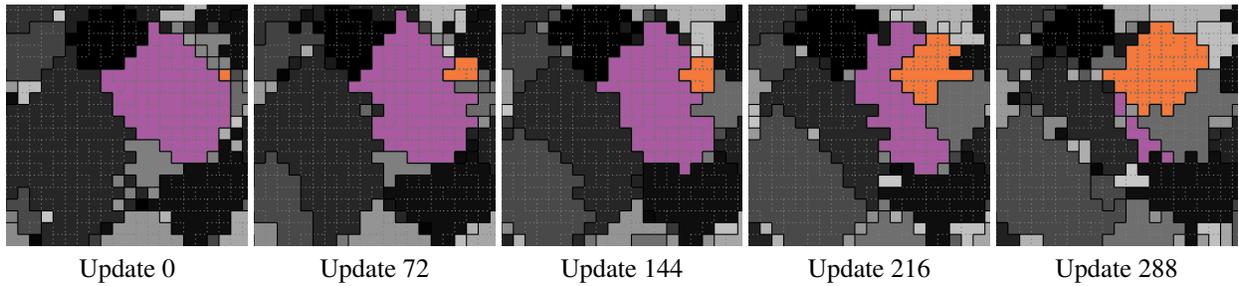
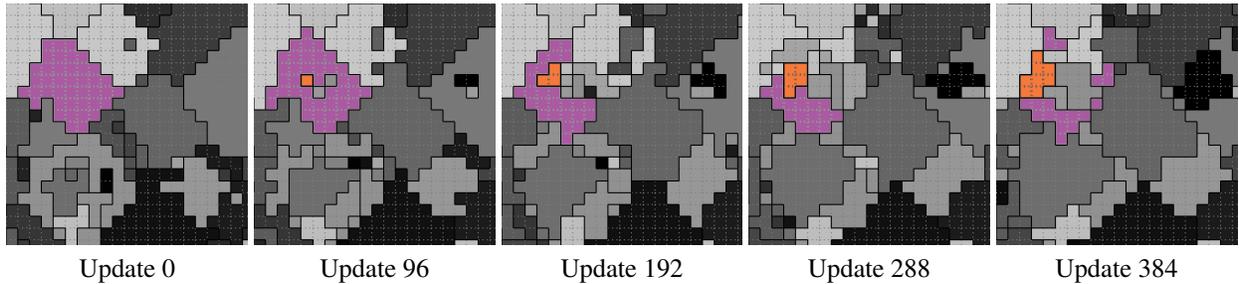

\begin{center}

\begin{subfigure}[b]{\textwidth}
\centering
\begin{minipage}[t]{0.18\textwidth}
\centering
\adjincludegraphics[width=\textwidth, trim={{.5\width} {.33\width} {.17\width} {.33\width}}, clip]{lifecycle/naive-paint/seed=1026+title=directional_channel_grayscale_viz+treat=resource-wave__channelsense-yes__nlev-two+update=1048576+_data_hathash_hash=fdd6f7a5210124bd+_script_fullcat_hash=602c0d0c070e9202+_source_hash=53a2252-clean+ext=}
Update 0
\end{minipage}
\begin{minipage}[t]{0.18\textwidth}
\centering
\adjincludegraphics[width=\textwidth, trim={{.5\width} {.33\width} {.17\width} {.33\width}}, clip]{lifecycle/naive-paint/seed=1026+title=directional_channel_grayscale_viz+treat=resource-wave__channelsense-yes__nlev-two+update=1048704+_data_hathash_hash=fdd6f7a5210124bd+_script_fullcat_hash=602c0d0c070e9202+_source_hash=53a2252-clean+ext=}
Update 128
\end{minipage}
\begin{minipage}[t]{0.18\textwidth}
\centering
\adjincludegraphics[width=\textwidth, trim={{.5\width} {.33\width} {.17\width} {.33\width}}, clip]{lifecycle/naive-paint/seed=1026+title=directional_channel_grayscale_viz+treat=resource-wave__channelsense-yes__nlev-two+update=1048832+_data_hathash_hash=fdd6f7a5210124bd+_script_fullcat_hash=602c0d0c070e9202+_source_hash=53a2252-clean+ext=}
Update 256
\end{minipage}
\begin{minipage}[t]{0.18\textwidth}
\centering
\adjincludegraphics[width=\textwidth, trim={{.5\width} {.33\width} {.17\width} {.33\width}}, clip]{lifecycle/naive-paint/seed=1026+title=directional_channel_grayscale_viz+treat=resource-wave__channelsense-yes__nlev-two+update=1048960+_data_hathash_hash=fdd6f7a5210124bd+_script_fullcat_hash=602c0d0c070e9202+_source_hash=53a2252-clean+ext=}
Update 384
\end{minipage}
\begin{minipage}[t]{0.18\textwidth}
\centering
\adjincludegraphics[width=\textwidth, trim={{.5\width} {.33\width} {.17\width} {.33\width}}, clip]{lifecycle/naive-paint/seed=1026+title=directional_channel_grayscale_viz+treat=resource-wave__channelsense-yes__nlev-two+update=1049088+_data_hathash_hash=fdd6f7a5210124bd+_script_fullcat_hash=602c0d0c070e9202+_source_hash=53a2252-clean+ext=}
Update 512
\end{minipage}
\caption{
Naive
(%
animation: \url{https://mmore500.com/hopto/x},
in-browser simulation: \url{https://mmore500.com/hopto/1}%
)
}
\label{fig:lifecycle-naive}
\end{subfigure}

\vspace{3ex}

\begin{subfigure}[b]{\textwidth}
\centering
\begin{minipage}[t]{0.18\textwidth}
\centering
\adjincludegraphics[width=\textwidth, trim={{.66\width} {.66\width} {.0\width} {.0\width}}, clip]{lifecycle/adjoin-paint/seed=1004+title=directional_channel_grayscale_viz+treat=resource-wave__channelsense-yes__nlev-two+update=1048896+_data_hathash_hash=51fd4923ae7e3bde+_script_fullcat_hash=602c0d0c070e9202+_source_hash=53a2252-clean+ext=}
Update 0
\end{minipage}
\begin{minipage}[t]{0.18\textwidth}
\centering
\adjincludegraphics[width=\textwidth, trim={{.66\width} {.66\width} {.0\width} {.0\width}}, clip]{lifecycle/adjoin-paint/seed=1004+title=directional_channel_grayscale_viz+treat=resource-wave__channelsense-yes__nlev-two+update=1048928+_data_hathash_hash=51fd4923ae7e3bde+_script_fullcat_hash=602c0d0c070e9202+_source_hash=53a2252-clean+ext=}
Update 32
\end{minipage}
\begin{minipage}[t]{0.18\textwidth}
\centering
\adjincludegraphics[width=\textwidth, trim={{.66\width} {.66\width} {.0\width} {.0\width}}, clip]{lifecycle/adjoin-paint/seed=1004+title=directional_channel_grayscale_viz+treat=resource-wave__channelsense-yes__nlev-two+update=1048960+_data_hathash_hash=51fd4923ae7e3bde+_script_fullcat_hash=602c0d0c070e9202+_source_hash=53a2252-clean+ext=}
Update 64
\end{minipage}
\begin{minipage}[t]{0.18\textwidth}
\centering
\adjincludegraphics[width=\textwidth, trim={{.66\width} {.66\width} {.0\width} {.0\width}}, clip]{lifecycle/adjoin-paint/seed=1004+title=directional_channel_grayscale_viz+treat=resource-wave__channelsense-yes__nlev-two+update=1049024+_data_hathash_hash=51fd4923ae7e3bde+_script_fullcat_hash=602c0d0c070e9202+_source_hash=53a2252-clean+ext=}
Update 128
\end{minipage}
\begin{minipage}[t]{0.18\textwidth}
\centering
\adjincludegraphics[width=\textwidth, trim={{.66\width} {.66\width} {.0\width} {.0\width}}, clip]{lifecycle/adjoin-paint/seed=1004+title=directional_channel_grayscale_viz+treat=resource-wave__channelsense-yes__nlev-two+update=1049408+_data_hathash_hash=51fd4923ae7e3bde+_script_fullcat_hash=602c0d0c070e9202+_source_hash=53a2252-clean+ext=}
Update 512
\end{minipage}
\caption{
Adjoin
(%
animation: \url{https://mmore500.com/hopto/y},
in-browser simulation: \url{https://mmore500.com/hopto/2}%
)
}
\label{fig:lifecycle-adjoin}
\end{subfigure}

\vspace{3ex}

\begin{subfigure}[b]{\textwidth}
\centering
\begin{minipage}[t]{0.18\textwidth}
\centering
\adjincludegraphics[width=\textwidth, trim={{.0\width} {.0\width} {.66\width} {.66\width}}, clip]{lifecycle/sweep-paint/seed=1023+title=directional_channel_grayscale_viz+treat=resource-wave__channelsense-yes__nlev-two+update=1048648+_data_hathash_hash=519a2d5a19f16020+_script_fullcat_hash=602c0d0c070e9202+_source_hash=53a2252-clean+ext=}
Update 0
\end{minipage}
\begin{minipage}[t]{0.18\textwidth}
\centering
\adjincludegraphics[width=\textwidth, trim={{.0\width} {.0\width} {.66\width} {.66\width}}, clip]{lifecycle/sweep-paint/seed=1023+title=directional_channel_grayscale_viz+treat=resource-wave__channelsense-yes__nlev-two+update=1048720+_data_hathash_hash=519a2d5a19f16020+_script_fullcat_hash=602c0d0c070e9202+_source_hash=53a2252-clean+ext=}
Update 72
\end{minipage}
\begin{minipage}[t]{0.18\textwidth}
\centering
\adjincludegraphics[width=\textwidth, trim={{.0\width} {.0\width} {.66\width} {.66\width}}, clip]{lifecycle/sweep-paint/seed=1023+title=directional_channel_grayscale_viz+treat=resource-wave__channelsense-yes__nlev-two+update=1048792+_data_hathash_hash=519a2d5a19f16020+_script_fullcat_hash=602c0d0c070e9202+_source_hash=53a2252-clean+ext=}
Update 144
\end{minipage}
\begin{minipage}[t]{0.18\textwidth}
\centering
\adjincludegraphics[width=\textwidth, trim={{.0\width} {.0\width} {.66\width} {.66\width}}, clip]{lifecycle/sweep-paint/seed=1023+title=directional_channel_grayscale_viz+treat=resource-wave__channelsense-yes__nlev-two+update=1048864+_data_hathash_hash=519a2d5a19f16020+_script_fullcat_hash=602c0d0c070e9202+_source_hash=53a2252-clean+ext=}
Update 216
\end{minipage}
\begin{minipage}[t]{0.18\textwidth}
\centering
\adjincludegraphics[width=\textwidth, trim={{.0\width} {.0\width} {.66\width} {.66\width}}, clip]{lifecycle/sweep-paint/seed=1023+title=directional_channel_grayscale_viz+treat=resource-wave__channelsense-yes__nlev-two+update=1048936+_data_hathash_hash=519a2d5a19f16020+_script_fullcat_hash=602c0d0c070e9202+_source_hash=53a2252-clean+ext=}
Update 288
\end{minipage}
\caption{
Sweep
(%
animation: \url{https://mmore500.com/hopto/z},
in-browser simulation: \url{https://mmore500.com/hopto/3}%
)
}
\label{fig:lifecycle-sweep}
\end{subfigure}

\vspace{3ex}

\begin{subfigure}[b]{\textwidth}
\centering
\begin{minipage}[t]{0.18\textwidth}
\centering
\adjincludegraphics[width=\textwidth, trim={{.0\width} {.66\width} {.66\width} {.0\width}}, clip]{lifecycle/burst-paint/seed=1034+title=directional_channel_grayscale_viz+treat=resource-wave__channelsense-yes__nlev-two+update=1048648+_data_hathash_hash=ca8ad21d3b30b939+_script_fullcat_hash=602c0d0c070e9202+_source_hash=53a2252-clean+ext=}
Update 0
\end{minipage}
\begin{minipage}[t]{0.18\textwidth}
\centering
\adjincludegraphics[width=\textwidth, trim={{.0\width} {.66\width} {.66\width} {.0\width}}, clip]{lifecycle/burst-paint/seed=1034+title=directional_channel_grayscale_viz+treat=resource-wave__channelsense-yes__nlev-two+update=1048744+_data_hathash_hash=ca8ad21d3b30b939+_script_fullcat_hash=602c0d0c070e9202+_source_hash=53a2252-clean+ext=}
Update 96
\end{minipage}
\begin{minipage}[t]{0.18\textwidth}
\centering
\adjincludegraphics[width=\textwidth, trim={{.0\width} {.66\width} {.66\width} {.0\width}}, clip]{lifecycle/burst-paint/seed=1034+title=directional_channel_grayscale_viz+treat=resource-wave__channelsense-yes__nlev-two+update=1048840+_data_hathash_hash=ca8ad21d3b30b939+_script_fullcat_hash=602c0d0c070e9202+_source_hash=53a2252-clean+ext=}
Update 192
\end{minipage}
\begin{minipage}[t]{0.18\textwidth}
\centering
\adjincludegraphics[width=\textwidth, trim={{.0\width} {.66\width} {.66\width} {.0\width}}, clip]{lifecycle/burst-paint/seed=1034+title=directional_channel_grayscale_viz+treat=resource-wave__channelsense-yes__nlev-two+update=1048936+_data_hathash_hash=ca8ad21d3b30b939+_script_fullcat_hash=602c0d0c070e9202+_source_hash=53a2252-clean+ext=}
Update 288
\end{minipage}
\begin{minipage}[t]{0.18\textwidth}
\centering
\adjincludegraphics[width=\textwidth, trim={{.0\width} {.66\width} {.66\width} {.0\width}}, clip]{lifecycle/burst-paint/seed=1034+title=directional_channel_grayscale_viz+treat=resource-wave__channelsense-yes__nlev-two+update=1049032+_data_hathash_hash=ca8ad21d3b30b939+_script_fullcat_hash=602c0d0c070e9202+_source_hash=53a2252-clean+ext=}
Update 384
\end{minipage}
\caption{
Burst
(%
animation: \url{https://mmore500.com/hopto/0},
in-browser simulation: \url{https://mmore500.com/hopto/4}%
)
}
\label{fig:lifecycle-burst}
\end{subfigure}

\caption{
Time lapse examples of qualitative life histories evolved under the Nested-Wave treatment.
Level-one groups are by differentiated by grayscale tone and separated by solid black borders.
Level-zero groups are by separated by dashed gray borders.
In each example, the focal parent level-one group is colored purple and the focal offspring group orange.
}
\label{fig:lifecycle}
\end{center}
\end{figure*}

%% file: fig/ko-interior_propagule.tex
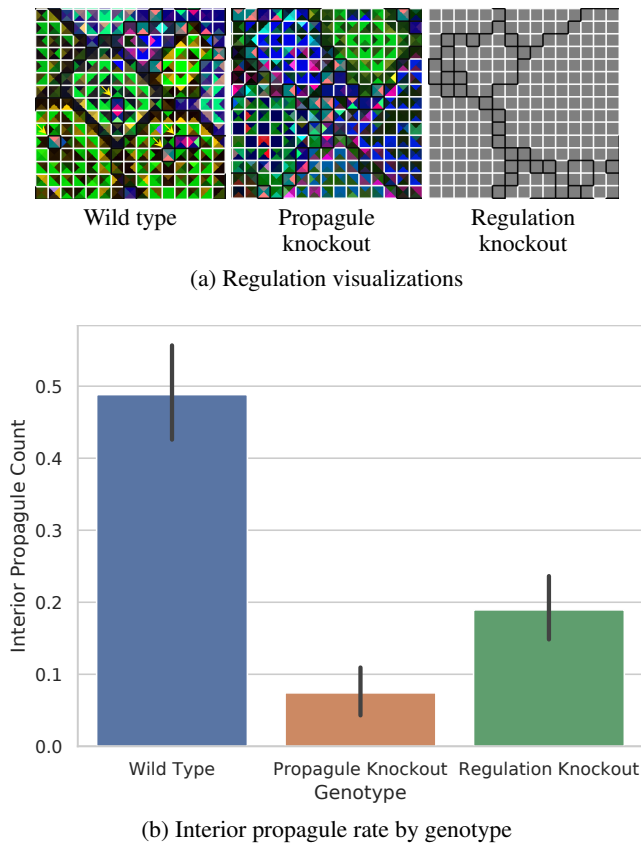
\begin{figure}[!htbp]
\begin{center}

\begin{subfigure}[b]{\linewidth}
\begin{center}

\begin{minipage}[t]{0.30\linewidth}
\centering
\vspace{0pt} 
\begin{tikzpicture}
\node[anchor=south west,inner sep=0] (image) at (0,0) { \adjincludegraphics[width=\linewidth, trim={{.25\width} {.20\width} {.5\width} {.55\width}}, clip]{knockout/interior_propagule/wildtype/seed=1+title=directional_regulator_viz+treat=resource-wave__channelsense-yes__nlev-two+update=8188+_data_hathash_hash=8b493febd79aad1f+_script_fullcat_hash=90718bb0c6ec4dbd+_source_hash=53a2252-clean+ext=}
};
\begin{scope}[x={(image.south east)},y={(image.north west)}]
  \draw [-stealth, yellow] (0.35,0.59) -- ++(0.05,-0.05);
  \draw [-stealth, yellow] (0.01,0.39) -- ++(0.05,-0.05);
  \draw [-stealth, yellow] (0.68,0.39) -- ++(0.05,-0.05);
  \draw [-stealth, yellow] (0.62,0.32) -- ++(0.05,-0.05);
\end{scope}
\end{tikzpicture}
\footnotesize Wild type
\end{minipage}
\begin{minipage}[t]{0.30\linewidth}
\centering
\vspace{0pt} 
\adjincludegraphics[width=\linewidth, trim={{.5\width} {.5\width} {.25\width} {.25\width}}, clip]{knockout/interior_propagule/propaguleknockout/seed=1+title=directional_regulator_viz+treat=resource-wave__channelsense-yes__nlev-two+update=8188+_data_hathash_hash=2b6711db47fb5887+_script_fullcat_hash=90718bb0c6ec4dbd+_source_hash=53a2252-clean+ext=}
\footnotesize Propagule knockout
\end{minipage}
\begin{minipage}[t]{0.30\linewidth}
\centering
\vspace{0pt} 
\adjincludegraphics[width=\linewidth, trim={{.5\width} {.5\width} {.25\width} {.25\width}}, clip]{knockout/interior_propagule/regulationknockout/seed=1+title=directional_regulator_viz+treat=resource-wave__channelsense-yes__nlev-two+update=8188+_data_hathash_hash=11ab5cdd47ed18c7+_script_fullcat_hash=90718bb0c6ec4dbd+_source_hash=53a2252-clean+ext=}
\footnotesize Regulation knockout
\end{minipage}

\caption{Regulation visualizations}
\label{fig:regulation_visualizations}

\end{center}
\end{subfigure}

\begin{minipage}[t]{\linewidth}
\centering
\vspace{0pt} 
\begin{subfigure}[b]{\linewidth}
\includegraphics[width=\linewidth]{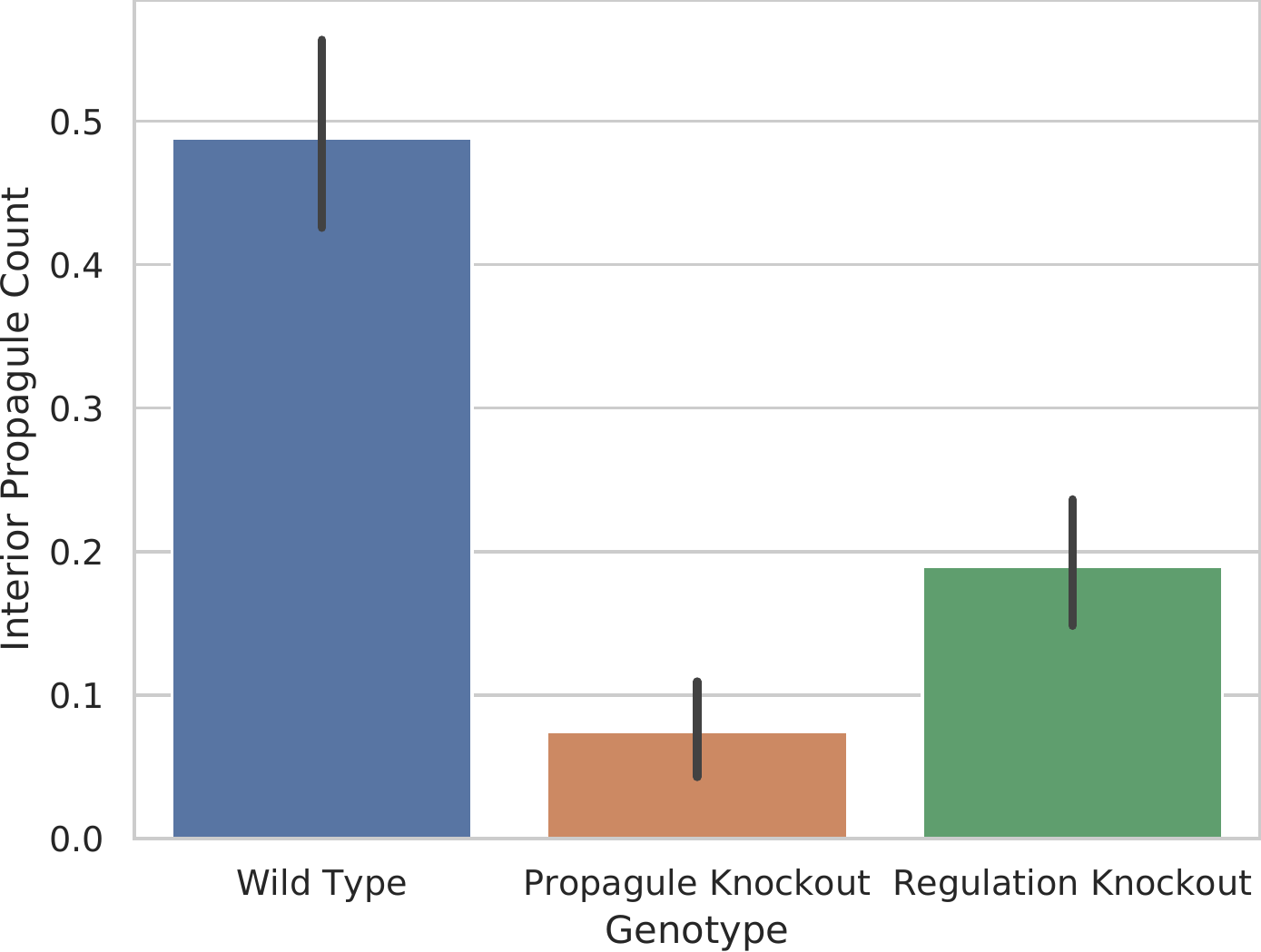}%
\caption{Interior propagule rate by genotype}
\label{fig:interior_propagule_rate}
\end{subfigure}
\end{minipage}%
\hspace*{\fill}

\caption{
Analysis of a wild type strain exhibiting a "burst" lifecycle evolved under the ``Nested-Wave'' treatment exhibiting interior propagule generation.
Figure \ref{fig:regulation_visualizations} depicts gene regulation at each of a cell's four directional SignalGP instances using a PCA mapping from regulatory state to three-dimensional RGB coordinates, calculated uniquely for each level-one same-channel signaling group.
Black borders divide outer registered-kin groups and white borders divide inner registered-kin groups.
Endogenous daughter groups annotated with yellow arrows.
Figure \ref{fig:interior_propagule_rate} compares the mean number of interior propagules observed per level-one same-channel signaling group.
Error bars indicate 95\% confidence.
View an animation of wild type gene regulation at \url{https://mmore500.com/hopto/t}.
View the wild type strain in a live in-browser simulation at \url{https://mmore500.com/hopto/g}.
}
\label{fig:ko-interior_propagule}
\end{center}
\end{figure}

%% file: fig/ko-intermessaging-sharing.tex
\begin{figure}[!htbp]
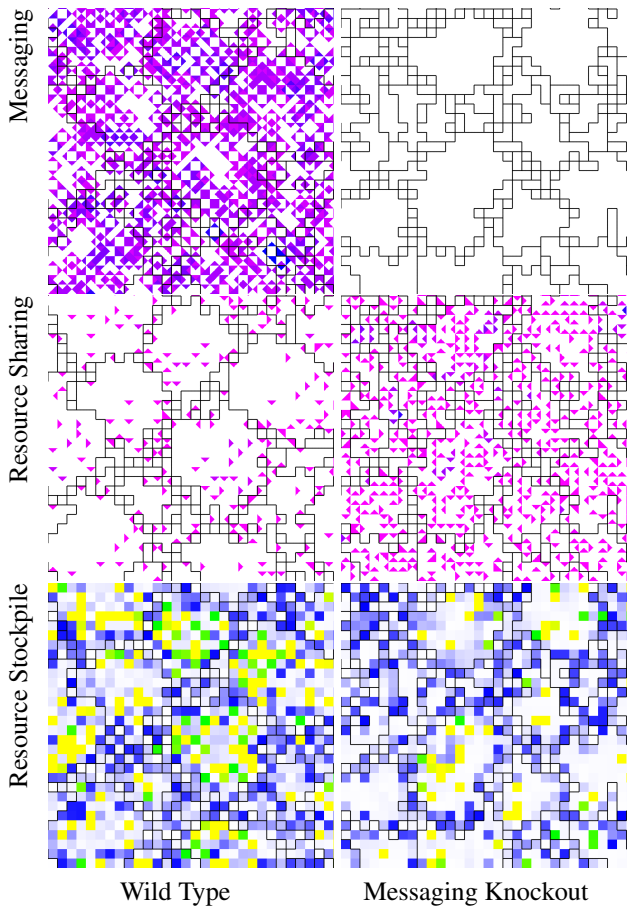

\begin{center}

\begin{minipage}[t]{\columnwidth}
\hspace*{\fill}%
\begin{minipage}[t]{0.05\columnwidth}
\vspace{0pt} 
\rotatebox{90}{Messaging}%
\end{minipage}%
\hfill
\begin{minipage}[t]{0.45\columnwidth}
\centering
\vspace{0pt} 
\adjincludegraphics[width=\textwidth, trim={{.0\width} {.0\width} {.5\width} {.5\width}}, clip]{knockout/intermessaging-sharing/wildtype/seed=1+title=directional_messaging_viz+treat=resource-wave__channelsense-yes__nlev-onebig+update=7172+_data_hathash_hash=f9e2a8ff33bf7745+_script_fullcat_hash=6b7e0389992dd616+_source_hash=53a2252-clean+ext=}%
\end{minipage}%
\hfill
\begin{minipage}[t]{0.45\columnwidth}
\centering
\vspace{0pt} 
\adjincludegraphics[width=\textwidth, trim={{.0\width} {.0\width} {.5\width} {.5\width}}, clip]{knockout/intermessaging-sharing/knockout/seed=1+title=directional_messaging_viz+treat=resource-wave__channelsense-yes__nlev-onebig+update=7172+_data_hathash_hash=ffdeb1c77dd012e1+_script_fullcat_hash=6b7e0389992dd616+_source_hash=53a2252-clean+ext=}%
\end{minipage}%
\hspace*{\fill}

\hspace*{\fill}%
\begin{minipage}[t]{0.05\columnwidth}
\vspace{0pt} 
\rotatebox{90}{Resource Sharing}%
\end{minipage}%
\hfill
\begin{minipage}[t]{0.45\columnwidth}
\centering
\vspace{0pt} 
\adjincludegraphics[width=\textwidth, trim={{.0\width} {.0\width} {.5\width} {.5\width}}, clip]{knockout/intermessaging-sharing/wildtype/seed=1+title=directional_sharing_viz+treat=resource-wave__channelsense-yes__nlev-onebig+update=7172+_data_hathash_hash=f9e2a8ff33bf7745+_script_fullcat_hash=3a1e851383e0ffd4+_source_hash=53a2252-clean+ext=}%
\end{minipage}%
\hfill
\begin{minipage}[t]{0.45\columnwidth}
\centering
\vspace{0pt} 
\adjincludegraphics[width=\textwidth, trim={{.0\width} {.0\width} {.5\width} {.5\width}}, clip]{knockout/intermessaging-sharing/knockout/seed=1+title=directional_sharing_viz+treat=resource-wave__channelsense-yes__nlev-onebig+update=7172+_data_hathash_hash=ffdeb1c77dd012e1+_script_fullcat_hash=3a1e851383e0ffd4+_source_hash=53a2252-clean+ext=}%
\end{minipage}%
\hspace*{\fill}

\hspace*{\fill}%
\begin{minipage}[t]{0.05\columnwidth}
\vspace{0pt} 
\rotatebox{90}{Resource Stockpile}%
\end{minipage}%
\hfill
\begin{minipage}[t]{0.45\columnwidth}
\centering
\vspace{0pt} 
\adjincludegraphics[width=\textwidth, trim={{.0\width} {.0\width} {.5\width} {.5\width}}, clip]{knockout/intermessaging-sharing/wildtype/seed=1+title=stockpile_viz+treat=resource-wave__channelsense-yes__nlev-onebig+update=7172+_data_hathash_hash=f9e2a8ff33bf7745+_script_fullcat_hash=4c8152cbf92e0da6+_source_hash=53a2252-clean+ext=}%
\end{minipage}%
\hfill
\begin{minipage}[t]{0.45\columnwidth}
\centering
\vspace{0pt} 
\adjincludegraphics[width=\textwidth, trim={{.0\width} {.0\width} {.5\width} {.5\width}}, clip]{knockout/intermessaging-sharing/knockout/seed=1+title=stockpile_viz+treat=resource-wave__channelsense-yes__nlev-onebig+update=7172+_data_hathash_hash=ffdeb1c77dd012e1+_script_fullcat_hash=4c8152cbf92e0da6+_source_hash=53a2252-clean+ext=}%
\end{minipage}%
\hspace*{\fill}

\vspace{1.0ex}

\hspace*{\fill}%
\begin{minipage}[t]{0.05\columnwidth}
\vspace{0pt} 
\end{minipage}%
\hfill
\begin{minipage}[t]{0.45\columnwidth}
\centering
\vspace{0pt} 
Wild Type
\end{minipage}%
\hfill
\begin{minipage}[t]{0.45\columnwidth}
\centering
\vspace{0pt} 
Messaging Knockout
\end{minipage}%
\hspace*{\fill}

\vspace{1.0ex}


\end{minipage}%
%

\caption{
Visualization of phenotypic traits of a wild type strain evolved under the ``Flat-Wave'' treatment and corresponding intercell messaging knockout strain.
In the messaging visualization, color coding represents the volume of incoming messages.
White represents no incoming messages and the magenta to blue gradient runs from one incoming message to the maximum observed incoming message traffic.
In the resource sharing visualization, color coding represents the amount of incoming resource.
In the resource stockpile visualization, white represents zero-resource stockpiles, blue represents stockpiles with just under enough resource to reproduce, green represents stockpiles with enough resource to reproduce, and yellow represents more than enough resource to reproduce.
Black borders divide level-zero same-channel signaling networks.
View an animation of the wild type strain at \url{https://mmore500.com/hopto/p}.
View the wild type strain in a live in-browser simulation at \url{https://mmore500.com/hopto/e}.
}
\label{fig:ko-intermessaging-sharing}
\end{center}
\end{figure}

%% file: fig/ko-stockpiletrigger-sharing.tex
\begin{figure}[!htbp]
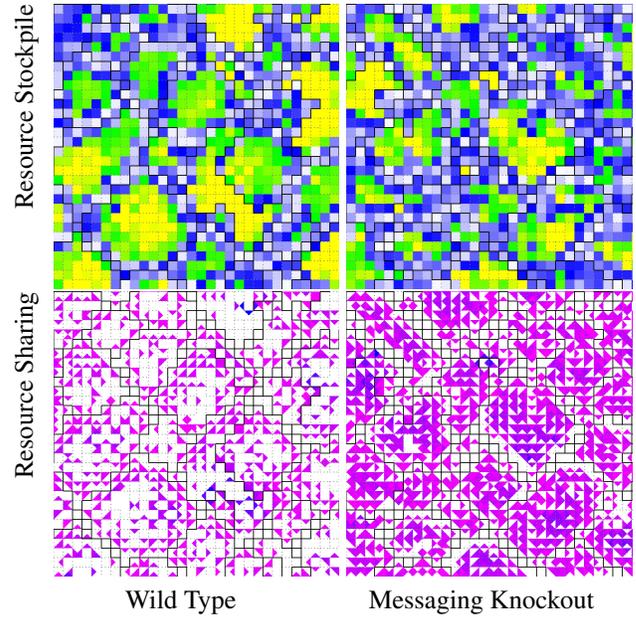

\begin{center}

\centering

\hspace*{\fill}%
\begin{minipage}[t]{0.05\columnwidth}
\vspace{0pt} 
\rotatebox{90}{Resource Stockpile}%
\end{minipage}%
\hfill
\begin{minipage}[t]{0.45\columnwidth}
\centering
\vspace{0pt} 
\adjincludegraphics[width=\textwidth, trim={{.0\width} {.0\width} {.5\width} {.5\width}}, clip]{knockout/stockpiletrigger-sharing/wildtype/seed=1+title=stockpile_viz+treat=resource-wave__channelsense-yes__nlev-two+update=7172+_data_hathash_hash=d856da4ae5863122+_script_fullcat_hash=4c8152cbf92e0da6+_source_hash=53a2252-clean+ext=}%
\end{minipage}%
\hfill
\begin{minipage}[t]{0.45\columnwidth}
\centering
\vspace{0pt} 
\adjincludegraphics[width=\textwidth, trim={{.0\width} {.0\width} {.5\width} {.5\width}}, clip]{knockout/stockpiletrigger-sharing/knockout/seed=1+title=stockpile_viz+treat=resource-wave__channelsense-yes__nlev-two+update=7172+_data_hathash_hash=6ab6ade50c5344bc+_script_fullcat_hash=4c8152cbf92e0da6+_source_hash=53a2252-clean+ext=}%
\end{minipage}%
\hspace*{\fill}

\hspace*{\fill}%
\begin{minipage}[t]{0.05\columnwidth}
\vspace{0pt} 
\rotatebox{90}{Resource Sharing}%
\end{minipage}%
\hfill
\begin{minipage}[t]{0.45\columnwidth}
\centering
\vspace{0pt} 
\adjincludegraphics[width=\textwidth, trim={{.0\width} {.0\width} {.5\width} {.5\width}}, clip]{knockout/stockpiletrigger-sharing/wildtype/seed=1+title=directional_sharing_viz+treat=resource-wave__channelsense-yes__nlev-two+update=7172+_data_hathash_hash=d856da4ae5863122+_script_fullcat_hash=3a1e851383e0ffd4+_source_hash=53a2252-clean+ext=}%
\end{minipage}%
\hfill
\begin{minipage}[t]{0.45\columnwidth}
\centering
\vspace{0pt} 
\adjincludegraphics[width=\textwidth, trim={{.0\width} {.0\width} {.5\width} {.5\width}}, clip]{knockout/stockpiletrigger-sharing/knockout/seed=1+title=directional_sharing_viz+treat=resource-wave__channelsense-yes__nlev-two+update=7172+_data_hathash_hash=6ab6ade50c5344bc+_script_fullcat_hash=3a1e851383e0ffd4+_source_hash=53a2252-clean+ext=}%
\end{minipage}%
\hspace*{\fill}

\vspace{1.0ex}

\hspace*{\fill}%
\begin{minipage}[t]{0.05\columnwidth}
\vspace{0pt} 
\end{minipage}%
\hfill
\begin{minipage}[t]{0.45\columnwidth}
\centering
\vspace{0pt} 
Wild Type
\end{minipage}%
\hfill
\begin{minipage}[t]{0.45\columnwidth}
\centering
\vspace{0pt} 
Messaging Knockout
\end{minipage}%
\hspace*{\fill}

\vspace{1.0ex}

\caption{
Visualization of phenotypic traits of a wild type strain evolved under the ``Nested-Wave''' treatment and corresponding resource-sensing knockout strain.
In the resource-sharing visualization, color coding represents the amount of incoming shared resource.
White represents no incoming messages and the magenta to blue gradient runs from one incoming message to the maximum observed incoming message traffic.
In the resource stockpile visualization, white represents zero-resource stockpiles, blue represents stockpiles with just under enough resource to reproduce, green represents stockpiles with enough resource to reproduce, and yellow represents more than enough resource to reproduce.
Black borders divide level-one same-channel signaling groups and white borders divide level-zero same-channel signaling groups.
View an animation of the wild type strain at \url{https://mmore500.com/hopto/s}.
View the wild type strain in a live in-browser simulation at \url{https://mmore500.com/hopto/h}.
}
\label{fig:ko-stockpiletrigger-sharing}
\end{center}
\end{figure}

%% file: fig/ko-apoptosis.tex
\begin{figure}[!htbp]
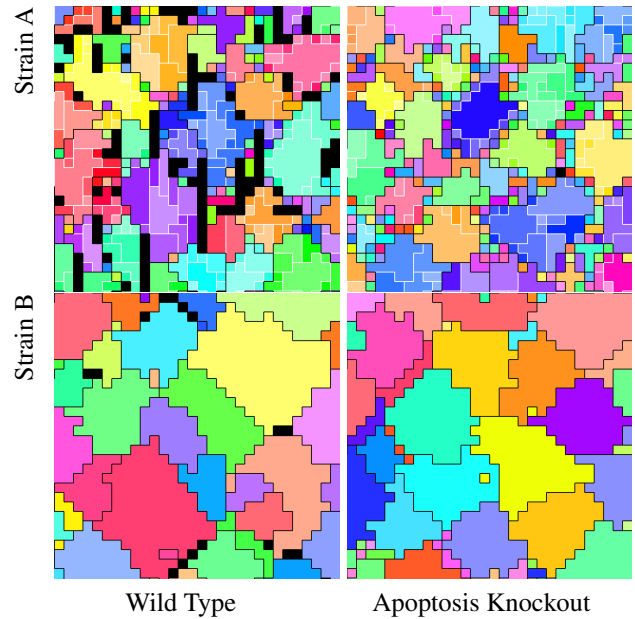

\begin{center}

\hspace*{\fill}%
\begin{minipage}[t]{0.05\columnwidth}
\vspace{0pt} 
\rotatebox{90}{Strain A}%
\end{minipage}%
\hfill
\begin{minipage}[t]{0.45\columnwidth}
\centering
\vspace{0pt} 
\adjincludegraphics[width=\textwidth, trim={{.5\width} {.5\width} {.0\width} {.0\width}}, clip]{knockout/apoptosis/wildtype/seed=1+title=channel_viz+treat=resource-even__channelsense-yes__nlev-two+update=262144+_data_hathash_hash=9b92a609c3309033+_script_fullcat_hash=7e789c981e3d0e4f+_source_hash=53a2252-clean+ext=}%
\end{minipage}%
\hfill
\begin{minipage}[t]{0.45\columnwidth}
\centering
\vspace{0pt} 
\adjincludegraphics[width=\textwidth, trim={{.5\width} {.5\width} {.0\width} {.0\width}}, clip]{knockout/apoptosis/knockout/seed=1+title=channel_viz+treat=resource-even__channelsense-yes__nlev-two+update=262144+_data_hathash_hash=900abeef45bb9133+_script_fullcat_hash=7e789c981e3d0e4f+_source_hash=53a2252-clean+ext=}%
\end{minipage}%
\hspace*{\fill}

\hspace*{\fill}%
\begin{minipage}[t]{0.05\columnwidth}
\vspace{0pt} 
\rotatebox{90}{Strain B}%
\hfill
\end{minipage}%
\hfill
\begin{minipage}[t]{0.45\columnwidth}
\centering
\vspace{0pt} 
\adjincludegraphics[width=\textwidth, trim={{.5\width} {.5\width} {.0\width} {.0\width}}, clip]{knockout/apoptosis/wildtype/seed=1+title=channel_viz+treat=resource-wave__channelsense-yes__nlev-onebig+update=8188+_data_hathash_hash=3465df2fce2dc5f4+_script_fullcat_hash=7e789c981e3d0e4f+_source_hash=53a2252-clean+ext=}
\end{minipage}%
\hfill
\begin{minipage}[t]{0.45\columnwidth}
\centering
\vspace{0pt} 
\adjincludegraphics[width=\textwidth, trim={{.5\width} {.5\width} {.0\width} {.0\width}}, clip]{knockout/apoptosis/knockout/seed=1+title=channel_viz+treat=resource-wave__channelsense-yes__nlev-onebig+update=8188+_data_hathash_hash=9c40470beee1c5b5+_script_fullcat_hash=7e789c981e3d0e4f+_source_hash=53a2252-clean+ext=}%
\end{minipage}%
\hspace*{\fill}

\vspace{1.0ex}

\hspace*{\fill}%
\begin{minipage}[t]{0.05\columnwidth}
\vspace{0pt} 
\end{minipage}%
\hfill
\begin{minipage}[t]{0.45\columnwidth}
\centering
\vspace{0pt} 
Wild Type
\end{minipage}%
\hfill
\begin{minipage}[t]{0.45\columnwidth}
\centering
\vspace{0pt} 
Apoptosis Knockout
\end{minipage}%
\hspace*{\fill}

\caption{
Comparison of wild type strains and corresponding apoptosis knockout strains.
In all visualizations, color hue denotes and black borders divide highest-level same-channel signaling networks.
In Replicate A visualizations, color saturation denotes and white borders divide level-zero same-channel signaling networks.
(Replicate B evolved under the flat treatment).
Black tiles are dead.
View an animation of wild type strain A at \url{https://mmore500.com/hopto/m}.
View an animation of wild type strain B at \url{https://mmore500.com/hopto/n}.
View wild type strain A in a live in-browser simulation at \url{https://mmore500.com/hopto/b}.
View wild type strain B in a live in-browser simulation at \url{https://mmore500.com/hopto/c}.
}
\label{fig:ko-apoptosis}
\end{center}
\end{figure}

%% file: tex/conclusion.tex
\section{Conclusion}


In this work, we selected for fraternal transitions individuality among digital organisms controlled by genetic programs.
Because --- unlike previous work --- we provided no experimentally-prescribed mechanism for collective reproduction, we observed the emergence of several distinct life histories. 
Evolved strategies exhibited intercellular communication, coordination, and differentiation.
These included endowment of offspring propagule groups, asymmetrical intra-group resource sharing, asymmetrical inter-group relationships, morphological patterning, gene-regulation mediated life cycles, and adaptive apoptosis.

Across treatments, we observed resource-sharing and reproductive cooperation among registered kin groups.
These outcomes arose even in treatments where registered kin groups lacked functional significance (i.e., resource was distributed evenly), suggesting that reliable kin recognition alone might be sufficient to observe aspects of fraternal collectivism evolve in systems where population members compete antagonistically for limited space or resources and spatial mixing is low.
In addition to their functional consequences, perhaps the role of physical mechanisms such as cell attachment simply as a kin recognition tool might merit consideration.

We plan to investigate mechanisms to evolve greater collective sophistication among agents.
The modular design of SignalGP lends itself to the possibility of exploring sexual recombination.
The spatial, distributed nature of our approach also affords a route to achieve large-scale digital multicellularity experiments consisting of millions, instead of thousands, of cells via high-performance parallel computing.
We are interested in the possibility of allowing cell groups to develop neural and vascular networks in such work.
We hypothesize that selective pressures related to intra-group coordination and inter-group conflict might spur developmental and structural infrastructure that could be co-opted to evolve agents proficient at unrelated tasks like navigation, game-playing, or reinforcement learning.

We are eager to undertake experiments investigating open questions pertaining to major evolutionary transitions such as the role of pre-existing phenotypic plasticity \citep{clune2007investigating, lalejini2016evolutionary}, pre-existing environmental interactions, pre-existing reproductive division of labor, and how transitions relate to increases in organizational \citep{goldsby2012task}, structural, and functional \citep{goldsby2014evolutionary} complexity.

%% file: tex/acknowledgements.tex
\section{Acknowledgements}

Thanks to members of the DEVOLAB, in particular Nathan Rizik for help implementing gene regulation features in SignalGP and Alexander Lalejini for help navigating the SignalGP programming API and graphically depicting SignalGP.
This research was supported in part by NSF grants DEB-1655715 and DBI-0939454 as well as by Michigan State University through the computational resources provided by the Institute for Cyber-Enabled Research.
This material is based upon work supported by the National Science Foundation Graduate Research Fellowship under Grant No. DGE-1424871.
Any opinions, findings, and conclusions or recommendations expressed in this material are those of the author(s) and do not necessarily reflect the views of the National Science Foundation.

%% file: tex/supplementary.tex
\section{Supplementary Material}

\subsection{Background} \label{sup:background}

In the domain of experimental evolution, William Ratcliff and collaborators imposed a selective pressure for hydrodynamic settling on Baker's yeast and observed, in response, the emergence of a multicellular snowflake morphology in which parent and daughter cells remained tethered \citep{ratcliff2014experimental}.
With this system, they showed that multicellular life history can arise from a single mutation and demonstrated that unicellular bottlenecking of lineages implicitly arises as an inherent geometric consequence of the snowflake morphology \citep{ratcliff2015origins}.
Under extreme settling selection pressure, they observed the emergence (and, encumbered by free riders, subsequent collapse) of altruistic behavior in which extracellular DNA and proteins released under elevated apoptosis rates scaffolds the formation of multi-group collectives \citep{gulli2019evolution}.
Separately, incomplete cell separation mediated by the same mutational pathway has also been observed to evolve in response to selection for extracellular sucrose digestion at low population densities \citep{koschwanez2013improved}.

A wealth of mathematical models spanning both reductive and agent-based approaches have been developed to describe evolution of multicellularity and cellular specialization \cite{hanschen2015evolutionary}.
Recent work by Staps and collaborators exemplifies the increasing mechanistic nuance of contemporary computational models.
They present a mechanistic, agent-based system in which cells evolve weights for a Boolean gene regulatory network with two inputs (representing environmental state), two hidden nodes (representing regulatory state), and two outputs that control reproduction rate and probability of dissociation from or association into a group (representing gene products).
Evolutionary runs reveal how ecological conditions, such as predation pressure, constraints on diffusion of nutrients/waste, and changing environmental conditions, influence multicellular evolutionary outcomes with respect to group size, group lifespan, group fertility, and cell fate \citep{staps2019emergence}.
Staps et al. identify further augmentation of the mechanistic capabilities of their agents --- particularly the capacity to establish explicit spatial spatial structure within groups and sense local state --- as a compelling target for future work.

Heather Goldsby and collaborators' deme-based work is illustrative of the artificial life approach, where focal structures and processes realize conceptual analogy to (but not necessarily direct representation of) biological reality.
In this string of studies, spatially-segregated pockets of cells (``demes'') compete for space in a fixed-size population of demes.
Individual cells are controlled by self-replicating Avida-style computer programs with special instructions that allow them to interact with their environment and with neighboring cells.
The free-form paradigm of the genetic programming substrate, theoretically capable of performing arbitrary computation, enables the evolution of agents exhibiting the advanced behavioral capacities proposed by Staps et al., albeit in a manner without direct mechanistic analogy to biological cells.
Two modes of reproduction are defined under the deme model: within-deme and deme-founding.
In the first, a cell copies itself into a neighboring toroidal tile within its deme.
In the second, a deme slot is cleared in the deme population then seeded with a single cell from the parent deme.
Cells grow freely within demes, but deme fecundity depends on the collective profile of computational tasks (e.g., logic functions) performed within the deme.
This setup mirrors the dynamics of biological multicellularity, in which cell proliferation may either grow an existing multicellular body or spawn a new multicellular body.
Notably, when task-switching costs are applied Goldsby et al. have observed the evolution of division of labor and extensive functional interdependence within demes \citep{goldsby2012task}.
When mutagenic side-effects are applied Goldsby et al. have observed the evolution of germ-soma differentiation \citep{goldsby2014evolutionary}.

\subsection{Resource Collection Process} \label{sup:resource_collection_process}

Resource appears at a single point then spreads outwards update-by-update in a diamond-shaped wave. The expanding wave halts at a predefined limit.
Cells must enter an ``activated'' state to harvest resource as it passes overhead.
The cell at the starting position of a resource wave is automatically activated, and will propagate the activation signal to neighboring cells on the same signaling channel.
The newly activated cells, in turn, activate their own neighbors registered to the same signaling channel.
Neighbors registered to other signaling channels do not activate.
Each cell, after sending the activation signal, enters a temporary quiescent state.
In this manner, cells sharing a signaling channel track and harvest an expanding resource wave.
The rate of resource collection for a cell is determined by the size and shape of of its same-channel signaling network;
small or fragmented same-channel signaling networks will frequently miss out on resource as it passes by.

Resource waves have a limited extent.
Cells that activate outside the extent of a resource wave collect no resource.
A long quiescent period ensures that erroneously activated cells miss several subsequent opportunities to collect resource and therefore will tend to collect resource at a slower rate.
In this manner, ``Goldilocks'' --- not to small and not too big --- signaling networks enjoy superior fitness.

Resource wave starting points (seeds) are tiled over the toroidal grid from a randomly chosen starting location such that the extents of the resource waves do not overlap.
All resource waves begin and proceed synchronously;
when they complete, the next resource waves are seeded.
This process provides efficient and spatially-uniform selection for ``Goldilocks'' same-channel signaling networks.

Cells control the size and shape of their same-channel signaling group through strategic reproduction.
Three choices are afforded: whether to reproduce at all, where among the four adjoining tiles of the toroidal grid to place their offspring, and whether the offspring should be registered to the parent's signaling channel or be given a random channel ID (in the range 1 to $2^{64} - 1$).
The probability of channel collision is miniscule: $60 \times 60 \times 2^{20}$ (the grid dimensions times the number of simulation updates) independent channel values will collide with probability less than $1 \times 10^{-9}$.
No guarantees are made about the uniqueness of a newly-generated channel ID, but chance collisions are rare.

In addition to ``signaling channel''-based resource collection, we provide a uniform inflow of $+0.0051$, sufficient for one reproduction approximately every thousand updates.

\subsection{Hierarchical Nesting} \label{sup:hierarchical_nesting}

Hierarchical levels are introduced into the system through multiple separate, but overlaid, instantiations of this resource wave/channel-signaling scheme.
We refer to each independent resource wave/channel-signaling system as a ``level.''
In some experimental treatments, we allowed two resource wave/channel-signaling levels, identified here as level one and level two.
On level one, resource waves extended a radius of two toroidal tiles.
On level two they extended a radius of six toroidal tiles.
On both levels, activated cells netted $+0.2$ resource from a resource wave, but did not collect any resource outside the extent of the resource wave.
Due to the different radii of resource waves on different levels, level one selects for small same-channel signaling networks and level two selects for large same-channel signaling networks.

Each cell contained a pair of separate channel IDs, the first for level one and the second for level two.
We kept these channel IDs hierarchically nested by constraining inheritance during reproduction.
Daughter cells could not inherit just the level-one channel ID, they could either
\begin{enumerate}
\item inherit both level-one and level-two channel ID,
\item inherit level-two channel ID but not level-one channel ID, or
\item inherit neither channel ID.
\end{enumerate}
Hierarchically nested channel IDs are analogous to a strict corporate organizational structure: all employees (i.e., cells) are members of one department (i.e., level-one channel network) and one corporation (i.e., level-two channel network) but no employee can be a member of two departments and no department can be a member of two corporations.
Figure \ref{fig:morphology-wt} depicts hierarchically nested channel states assumed by an evolved strain.

An evolutionary transition in individuality can readily be evaluated within the DISHTINY framework with respect to same-channel network groups.
In addition to a potentially functionally cooperative relationship, shared channel IDs --- which may only systematically arise through inheritance --- imply a close hereditary relationship.
Because new channel IDs arise first in a single cell, same-channel signaling networks are reproductively bottlenecked analogously to a "Staying Together" life cycle (rather than a "Coming Together" life cycle) \cite{staps2019emergence}.
This precludes chimeric groups, except for mutations arising from somatic reproduction and rare cases of channel ID collision.

To recognize an evolutionary transition in individuality, we can evaluate
\begin{enumerate}
\item whether cells with the same channel ID cooperate altruistically by assessing, for example, resource sharing, and
\item whether division of reproductive labor arises by assessing whether interior cells cede reproduction to those at the periphery.
\end{enumerate}
If cells sharing the same level-one channel fulfill these conditions, we would suppose that a first-level transition in individuality had occurred.
Likewise, if cells sharing the sharing the same level-two channel fulfill these conditions, we would suppose that a second-level transition in individuality had occurred.
Further, we can screen for the evolution of complex multicellularity by assessing cell-cell messaging, regulatory patterning, and functional differentiation between cells within the a same-channel signaling network \cite{knoll2011multiple}.

\subsection{Channel Group Life Cycle} \label{sup:channel_group_life_cycle}

Mature same-channel resource collecting groups enjoy a considerable advantage over fledging propagules.
Because of the isometric scaling relationship between surface area and perimeter, cooperative same-channel resource collecting groups can marshal more resource at their periphery.
In addition, because of their greater surface area, mature same-channel resource collecting groups are able to seed resource-wave events and collect resource at a higher per-cell rate.

In order to ensure channel group turnover and facilitate channel group propagation, we impose a timed phase-out of somatic reproduction and resource wave harvests.
For each cell, we track a channel generation counter at each resource wave level.
At the genesis of a new channel group, these counters are set to zero.
Daughter cells that expand a channel group's soma are initialized to a counter value one greater than their parent.
Additionally, all channel generation counters are incremented every 512 updates to ensure that soma ages even in the absence of reproduction.
When a cell's channel generation counter reaches 1.5 times the wave radius of its level, it can no longer produce somatic daughter cells.
Then, after two additional counter steps, cells lose their ability to seed resource wave events and collect resource.
Thus, as channel groups age over time, their constituent cells lose the ability regenerate somatic tissue and then, soon after, to collect resource.
To prevent complete stagnation in the case where all cells' channel generation counters expire we provide a uniform inflow of $+0.0051$, sufficient for one reproduction approximately every thousand updates.

Interaction between nested channel groups produces a notable selective byproduct.
Because smaller, level-one channel groups tend to have intrinsically shorter lifespans, in order to achieve the full potential productive somatic lifespan of a larger, level-two channel group its constituent small channel groups must be intermittently regenerated.
Otherwise, the soma's capacity to seed resource-wave events and to collect resource will be prematurely lost once its constituent smaller, level-one channel groups expire.

This aging scheme's design ultimately stems from a desire
\begin{enumerate}
\item to facilitate evolution through regular turnover of emergent individuals and
\item to scaffold workable propagation for primitive cellular strategies while furnishing opportunities for more sophisticated adaptations to the imposed life cycle constraints.
\end{enumerate}
However, in some sense the aging scheme is heavy-handed, in effect enforcing rather than enabling a birth-death life cycle.
The evolutionary basis of aging and mortality --- in particular, the possibility of intrinsic evolutionary adaptations promoting these phenomena in addition to extrinsic factors  --- remains an active topic of scientific discussion \cite{baig2014evolution}.
In future work, we are interested in evaluating the outcomes of relaxing constraints of this aging scheme under different evolutionary conditions (such as cosmic ray mutations or irregular population structure) in light of theory attributing mortality and aging to evolvability, mutational accumulation, and costly somatic maintenance.

\subsection{Cell-Level Organisms} \label{sup:cell_level_organisms}

\include{fig/signalgp-dishtinygp}

SignalGP programs are collections of independent procedural functions, each equipped with a bit-string tag \cite{lalejini2018evolving}.
A function is triggered by a signal with affinity that maximally and sufficiently matches its tag.
(A binding threshold of 0.1 was used in these experiments).
Signals may be generated by the environment, received as messages from other agents, or triggered internally by function execution.
Signals, and the ensuing chains of procedural execution they give rise to, are processed pseudo-concurrently by 24 virtual CPU cores.
Figure \ref{fig:signalgp-cartoon} schematically depicts a single SignalGP instance.

In this work, we introduce a regulatory extension to the SignalGP system.
During runtime, instructions may increase or decrease each tagged function's intrinsic tendency to match with --- and activate in response to --- tagged queries.
Intrinsic tag-to-tag match distances $m$ are modulated by a regulator value $r$ (baseline, 1.0) to become $r + r \times m$.
This scheme allows a function to be upregulated such that every query activates that function (e.g., $r = 0$) or no query activates that function (e.g., $r = \texttt{inf}$).
These regulation settings are heritable during reproduction but automatically decay after a number of updates determined when they are set.

To allow cells to protect themselves form potentially antagonistic interactions with their neighbors, we filter intercellular messages through a tag-matching membrane.
At runtime, cells can embed tags in this membrane that either admit or repel incoming messages.
Messages that do not match with a membrane tag are repelled.
A message, for example, that would activate a SignalGP function containing an apoptosis instruction could be rejected while other messages are accepted.
Tags embedded in this membrane automatically decay and may also be regulated.
We also filter messages between hardware instances within the same cell through a tag-matching membrane, but the default behavior for messages with unmatched tags is admission rather than rejection.

Previous work evolving digital organisms in grid-based problem domains has relied on a single computational instance which designates a direction to act in via an explicit cardinal ``facing'' state or output \cite{goldsby2014evolutionary, goldsby2018serendipitous, grabowski2010early, biswas2014causes, lalejini2018evolving}.
Under this paradigm, a large portion of genotype space encodes behaviors that are intrinsically asymmetrical with respect to absolute or relative (depending on implementation) cardinal direction.
However, in grid-based tasks, directional phenotypic symmetry is generally advantageous.
That is --- in the absence of a polarizing external stimulus --- successful agents generally behave uniformly with respect to each cardinal direction of the grid.
In this work, each cell employs four instances of SignalGP hardware: one ``facing'' each cardinal direction.
These computational instances all execute the same SignalGP program but are otherwise decoupled and may follow independent chains of execution and develop independent regulatory states.
Instances within a cell execute round robin step-by-step in an order that is randomly drawn at the outset of each update.

Genetic encodings that exploit problem-domain symmetries are known to promote evolvability and --- ultimately --- evolved solution quality \cite{clune2011performance, cheney2014unshackling}.
We submit that this directional hardware replication protocol likely increases the fraction of genotype space that encodes cardinally-symmetric phenotypes and therefore better facilitates the evolution of high-fitness phenotypes.
In further work, we look forward to exploring the evolvability and solution quality implications of this new approach.

The single SignalGP program that is mirrored across the cell's computational instances represents the cell's genome.
Mutation, with standard SignalGP mutation parameters as in \cite{lalejini2018evolving}, is applied to 1\% of daughter cells at birth.
In addition, genomes encode the bitstrings associated with environmental events.
These bitstrings evolve at a per-bit mutation rate equivalent to the bitstring labels of SignalGP functions.

Instances within a cell may send intracellular messages to one another or intercellular messages to a neighboring cell.
Intercellular messages are received by the SignalGP instance that faces the sending cell.
Figure \ref{fig:dishtinygp-cartoon} schematically depicts the configuration of the four SignalGP instances that constitute a single DISHTINY cell as well as the instances of neighboring cells that receive extracellular messages from the focal cell.

\subsection{Standard SignalGP Instruction Library} \label{sup:standard_instruction_library}

The default SignalGP instruction set defines a number of generic arithmetic, logic, utility, and program flow instructions \cite{lalejini2018evolving}.
We include these instructions in our experiment's instruction library.

To counteract crowding of the mutational landscape by the volume of custom instructions provided, a second identical copy of each standard SignalGP instruction was included in the library.

\begin{itemize}
\item \textbf{Increment}
Increment value in a designated register.
\item \textbf{Decrement}
Decrement value in a designated register.
\item \textbf{Not}
Logically toggle value in a designated register.
\item \textbf{Add}
Add values from two designated registers into a third designated register.
\item \textbf{Subtract}
Subtract values from two designated registers into a third designated register.
\item \textbf{Subtract}
Subtract values from two designated registers into a third designated register.
\item \textbf{Multiply}
Multiply values from two designated registers into a third designated register.
\item \textbf{Divide}
Divide values from two designated registers into a third designated register.
\item \textbf{Modulus}
Calculate the modulus from two designated registers and place result into a third designated register.
\item \textbf{Test Equal}
Compare values in two designated registers and place equality result into a third designated register.
\item \textbf{Test Non-equality}
Compare values in two designated registers and place opposite equality result into a third designated register.
\item \textbf{Test Less}
Compare values in two designated registers and place less-than result into a third designated register.
\item \textbf{If}
If a designated register is non-zero, proceed.
Otherwise, skip block.
\item \textbf{While}
While a designated register is non-zero, loop over a program block.
Otherwise, skip block.
\item \textbf{Countdown}
While a designated register is non-zero, loop over a program block and decrement the value in the designated register.
Otherwise, skip block.
\item \textbf{Close}
If a preceding program block is, close it.
\item \textbf{Break}
Break to the end of the current program block.
\item \textbf{Call}
Call the SignalGP program module that best matches instruction's affinity.
\item \textbf{Return}
If possible, return from the current function.
\item \textbf{Set Memory}
Set a designated register's value to hard-coded memory value.
\item \textbf{Set True}
Set a designated register's value to true (1.0).
\item \textbf{Copy Memory}
Copy the value of a designated register to a second designated register.
\item \textbf{Swap Memory}
Swap the values of two designated registers.
\item \textbf{Input}
Copy a designated element of input memory into a designated register.
\item \textbf{Output}
Copy to a designated element of output memory from a designated register.
\item \textbf{Commit}
Copy a designated register into a designated element of global memory.
\item \textbf{Pull}
Copy a designated element of global memory into a designated register.
\item \textbf{Fork}
Fork a new thread with the SignalGP program module that best matches the instruction's affinity.
\item \textbf{Terminate}
Terminate the current thread.
\item \textbf{Nop}
No operation.
\item \textbf{RNG Draw}
Draw a random value between 0.0 and 1.0 from random number generator and store result in a register.
\item \textbf{Set Regulator}
Set the program module regulator that best matches (without regulation) the instruction's affinity to the value of a designated register.
\item \textbf{Set Own Regulator}
Set the program module regulator of the currently-executing program module to value of a designated register.
\item \textbf{Adjust Regulator}
Adjust the program module regulator of the program module that best matches (without regulation) the instruction's affinity a designated fraction toward a designated register's value.
\item \textbf{Adjust Own Regulator}
Adjust the program module regulator of the currently-executing program module a designated fraction toward a designated register's value.
\item \textbf{Extend Regulator}
Adjust the program module regulator decay timer of the program module that best matches (without regulation) the instruction's by a designated register's value.
\item \textbf{Sense Regulator}
Copy the program module regulator value of the program module that best matches (without regulation) the instruction's affinity into a designated register.
\item \textbf{Sense Own Regulator}
Copy the program module regulator value of currently-executing program module into a designated register.
\end{itemize}

\subsection{Custom Instruction Library} \label{sup:custom_instruction_library}

We define a number of custom instructions to allow evolving programs to sense and interact with their environments, including
\begin{itemize}
\item reproduction,
\item resource sharing,
\item channel ID sensing,
\item apoptosis,
\item intracellular messaging, and
\item intercellular messaging.
\end{itemize}

We provide an listing of our experiment's instruction library below.

Instructions that involve an extracellular neighbor default to the cell that the executing SignalGP instance is facing.
To ensure a founding crop of viable individuals, apoptosis and program flow instructions in the initial randomly-generated population were replaced with no-op instructions.
However, these instructions were allowed to mutate in to genomes freely once evolutionary runs began.

\begin{itemize}
\item \textbf{Send Intracellular Message}
Send a message to a single other SignalGP instance within the cell specified by a designated register's value.
\item \textbf{Broadcast Intracellular Message}
Send a message to all SignalGP instances within the cell, excluding self.
\item \textbf{Put Internal Membrane Gatekeeper}
Place a tag in the internal membrane that, depending on insertion order, admits or blocks incoming internal messages it matches with.
\item \textbf{Send Intercellular Message}
Send a message to a single cellular neighbor.
\item \textbf{Broadcast Intercellular Message}
Send a message to all cellular neighbors.
\item \textbf{Put External Membrane Gatekeeper}
Place a tag in the external membrane that, depending on insertion order, admits or blocks incoming external messages it matches with.
\item \textbf{Set External Membrane Regulator}
Set the regulation of the gatekeeper in the external membrane that best matches (without regulation) the instruction's affinity to the value of a designated register.
\item \textbf{Adjust External Membrane Regulator}
Adjust the regulation of the gatekeeper in the external membrane that best matches (without regulation) the instruction's affinity a designated fraction toward a designated register's value.
\item \textbf{Sense External Membrane Regulator}
Copy the regulation value of the program module that best matches (without regulation) the instruction's affinity into a designated register.
\item \textbf{Activate Intercellular Inbox}
Mark the intercellular inbox to accept messages.
At cell birth, the inbox is deactivated.
\item \textbf{Deactivate Intercellular Inbox}
Mark the intercellular inbox to decline messages.
\item \textbf{Share Resource}
Send a proportion of the cell's stockpiled resource to a neighboring cell.
One instruction defaults to sending a large proportion of available resource (50\%) to the neighboring cell.
A second instruction defaults to sending a small proportion of available resource (5\%) to the neighboring cell.
The proportion of available resource can be adjusted by a register-based argument.
\item \textbf{Set Stockpile Sharing Reserve}
Designate a quantity of stockpiled resource as ineligible for sharing.
The amount may be modified by a register-based argument.
\item \textbf{Clear Stockpile Sharing Reserve}
Designate all stockpiled resource as eligble for sharing.
\item \textbf{Restrict Outgoing Shared Resource}
Reduce outgoing sharing efficacy.
Unsent resource is retained by the sending cell (with no resource lost).
The fraction reduced is determined by a register-based argument.
\item \textbf{Restrict Incoming Shared Resource}
Reduce incoming sharing efficacy.
Declined resource is retained by the sending cell (with no resource lost).
The fraction reduced is determined by a register-based argument.
\item \textbf{Reproduce}
Attempt to spawn a child cell in a particular direction, paid for out of the parent cell's resource stockpile.
If sufficient resource is not available in the cell's stockpile, no resource is action is taken.
Variants of this instruction are defined for each channel ID inheritance level: from endowing the daughter cell with the parental channel IDs across all levels, to endowing the daughter cell with a new level-one channel ID but the parent's level-two channel ID, to endowing the daughter cell with all-new channel IDs.
If a channel generation counter limit has been reached, reproduction is simply attempted at the next highest level; even with channel generation counters maxed out, cells may generate offspring with all-new channel IDs.
\item \textbf{Pause Reproduction}
Pause cellular reproduction in a single direction for the remainder of the current update and for the entire next update.
Variants of this instruction pause reproduction at a certain wave/channel-signaling level or across all channel ID inheritance levels.
\item \textbf{Set Stockpile Reproduction Reserve}
Designate a quantity of stockpiled resource as ineligible for use to reproduce.
The amount may be modified by a register-based argument.
\item \textbf{Clear Stockpile Sharing Reserve}
Designate all stockpiled resource as eligible for use to reproduce.
\item \textbf{Apoptosis}
The cell is killed at the end of the current update.
\item \textbf{Designate/Revoke Heir} A dying cell's own stockpile is split evenly among neighboring cells that are designated at the time of death.
On apoptosis, 50\% of the reproduction cost to establish a cell is also split between designated neighboring cells.
These instructions mark or un-mark a neighbor as a heir.
\item \textbf{Increase Channel Generation Counter}
Increases the cell's channel generation counter.
The amount the cell's generation counter is increased by can be adjusted by register-based argument.
\item \textbf{Query Own Stockpile}
Sets a designated register to the amount of resource present in the cell's stockpile.
\item \textbf{Query Own Channel Generation Counter}
This instruction sets a designated register to the value of the cell's channel generation counter.
A variant of this instruction is provided for each wave/channel-signaling level.
\item \textbf{Query ``Is Neighbor Live?''}
This instruction sets a designated register to 1 if the neighboring tile contains a live cell and 0 otherwise.
\item \textbf{Query ``Is Neighbor My Cellular Child?''}
This instruction sets a designated register to 1 if the neighboring cell is the daughter of the querying cell and 0 otherwise.
\item \textbf{Query ``Is Neighbor My Cellular Parent?''}
This instruction sets a designated register to 1 if the neighboring cell is the parent of the querying cell and 0 otherwise.
\item \textbf{Query ``Does Neighbor's Channel ID Match Mine?''}
This instruction sets a designated register to 1 if the neighboring cell has the same channel ID as the querying cell and 0 otherwise.
A variant of this instruction is provided for each wave/channel-signaling level.
\item \textbf{Query ``Does Neighbor's Channel ID Descend From Mine?''}
This instruction sets a designated register to 1 if the neighboring cell's highest-level channel ID is different from the querying cell's highest-level channel ID, but is descended from the querying cell's channel ID via an explicit propagule-generating reproduction call.
This instruction allows a querying cell to sense whether its neighbor is a member of a same-channel group that is a propagule of the querying cell's same-channel group.
\item \textbf{Query ``Does My Channel ID Descend From Neighbor's?''}
This instruction sets a designated register to 1 if the querying cell's highest-level channel ID is different from the neighboring cell's highest-level channel ID, but is descended from the neighboring cell's channel ID  via an explicit propagule-generating reproduction call.
This instruction allows a querying cell to sense whether it is a member of a same-channel group that is a propagule of the neighboring cell's same-channel group.
\item \textbf{Query ``Is Neighbor Poorer?''}
This instruction sets a designated register to 1 if the querying cell's resource stockpile is larger than the neighboring cell's.
\item \textbf{Query ``Is Neighbor Older?''}
This instruction sets a designated register to 1 if the querying cell's cell age is less than the neighboring cell's.
\item \textbf{Query ``Is Neighbor Expired?''}
This instruction sets a designated register to 1 if a neighboring cell's channel generation counter has exceeded the expiration threshold.
\item \textbf{Query Neighbor's Channel ID}
This instruction sets a designated register to the neighbor's channel ID.
A variant of this instruction is provided for each wave/channel-signaling level.
\item \textbf{Query Neighbor's Stockpile}
This instruction sets a designated register to the amount of resource present in the neighbor's stockpile.
\end{itemize}

\subsection{Environmental Cue Library} \label{sup:environmental_cue_library}

Event-driven sensing has been shown to enable evolution of SignalGP programs that more successfully react to  environmental state \cite{lalejini2018evolving}, so we supplement our instruction-based sensors with event-based input.
Every eight updates, a subset of environmental events are triggered on each SignalGP hardware based on current local environmental conditions.
The activating affinity of each event is genetically-encoded as part of the program currently executing on the hardware.
We provide a listing of our experiment's event library in supplementary material \ref{sup:environmental_cue_library}.

\begin{itemize}
\item \textbf{On Update}
This event is triggered every eight updates.
\item \textbf{Just Born}
This event is triggered once after a cell is born.
\item \textbf{Richer Neighbor}
This event is triggered if a neighbor cell has more stockpiled resource than the focal cell.
\item \textbf{Poorer Neighbor}
This event is triggered if a neighbor cell has less stockpiled resource than the focal cell.
\item \textbf{Facing Cellular Child}
This event is triggered if the SignalGP instance is facing a neighboring cell that is the querying cell's daughter.
\item \textbf{Facing Cellular Parent}
This event is triggered if the SignalGP instance is facing a neighboring cell that is the querying cell's parent.
\item \textbf{Neighbor's Channel ID Descends From Mine}
This event is triggered if the neighboring cell's highest-level channel ID is different from the querying cell's highest-level channel ID, but is descended from the querying cell's channel ID via an explicit propagule-generating reproduction call.
This event allows a querying cell to sense whether its neighbor is a member of a same-channel group that is a propagule of the querying cell's same-channel group.
\item \textbf{My Channel ID Descends From Neighbor's}
This event is triggered if the focal cell's highest-level channel ID is different from the neighboring cell's highest-level channel ID, but is descended from the neighboring cell's channel ID via an explicit propagule-generating reproduction call.
This event allows a neighboring cell to sense whether its neighbor is a member of a same-channel group that is a propagule of the neighboring cell's same-channel group.
\item \textbf{Neighbor's Channel ID Matches Mine}
This event is triggered if a SignalGP instance is facing a neighbor cell that shares its channel ID.
A different event is provided for each resource wave/channel-signaling level.
\item \textbf{Neighbor's Channel ID Does Not Match Mine}
This event is triggered if a SignalGP instance is facing a neighbor cell that does not share its channel ID.
A different event is provided for each resource wave/channel-signaling level.
\item \textbf{Channel Generation Counter Is Unexpired}
This event is triggered if a SignalGP instance's cell's channel generation counter has not yet reached the expiration threshold.
A different event is provided for each resource wave/channel-signaling level.
\item \textbf{Channel Generation Counter Is Expiring}
This event is triggered if a SignalGP instance's cell's channel generation counter has not yet reached the threshold where somatic propagation capacity, but not resource accumulation capacity, is lost.
A different event is provided for each resource wave/channel-signaling level.
\item \textbf{Channel Generation Counter Is Expired}
This event is triggered if a SignalGP instance's cell's channel generation counter has not yet reached the threshold where both somatic propagation capacity and resource accumulation capacity are lost.
A different event is provided for each resource wave/channel-signaling level.
\item \textbf{No-reward Resource Activation}
This event is triggered if a SignalGP instance's cell experiences a resource collection activation where no resource reward is achieved (e.g., the cell lies extent of the resource wave).
A different event is provided for each resource wave/channel-signaling level.
\end{itemize}

\subsection{Treatments} \label{sup:treatments}

\input{fig/productivity.tex}

\input{fig/systematics.tex}

In this work, we screened replicates conducted under combinations of two experimental conditions:
\begin{enumerate}
\item flat versus nested hierarchical resource wave/channel-signaling levels and
\item cooperative versus independent resource collection.
\end{enumerate}

The first experimental manipulation explores the effects of hierarchical nesting of kin-sensing and/or functional cooperation.
The second manipulation explores the effects of functional cooperation.

To enact the first manipulation, we compared the nested hierarchical resource wave/channel-signaling scheme described above with a single-level scheme with waves extending six toroidal tiles.
We also increased the resource wave reward to $+0.6$ to approximately match the observed resource inflow rate of the nested scheme.
To enact the second manipulation, we removed the resource wave reward and increased the uniform resource inflow rate to $+0.0175$ in order to approximately match the net inflow rate under the dual-level wave-based scheme.
Table \ref{tab:productivity} reports productivity observed under these different conditions.

We mix and match these experimental manipulations in three treatments:
\begin{enumerate}
\item one level with even resource (``Flat-Even''; in-browser simulation \url{https://mmore500.com/hopto/i}),
\item one level with wave-based resource (``Flat-Wave''; in-browser simulation \url{https://mmore500.com/hopto/j}),
\item two levels with even resource (``Nested-Even''; in-browser simulation \url{https://mmore500.com/hopto/k}), and
\item two levels with wave-based resource (``Nested-Wave''; in-browser simulation \url{https://mmore500.com/hopto/l}).
\end{enumerate}

We ran 40 replicates under each treatment condition.
Replicates were seeded with randomly generated SignalGP programs.
To conserve disk space, we divided evolutionary runs into 262144 ($2^{18}$) update epochs and collected data in 8096 ($2^{13}$) update snapshots between epochs.
All replicates ran at least one full epoch, and all comparisons between or within treatments are conducted at this time point.
However, most replicates (156/160) were able to run to four epochs during available compute time.
We screened for and conducted case studies at the latest available data for each replicate.
All reported case studies happen to be drawn from runs that completed 4 epochs of evolution.
Table \ref{tab:systematics} reports the systematics outcomes observed under each treatment at epoch 1 and at epoch 4.

All experiments took place a traditional 60-by-60 toroidal grid, supporting a population of at most 3600 individual cells.

\subsection{Competition Experiments and Phenotype Assays} \label{sup:competition_assays}

We performed further experiments to develop case studies of evolved strains we manually screened from our evolutionary runs.
In these experiments, the most-abundant genotype was harvested from the end-state of evolutionary runs as the wild type strain.
We collected epigenetic state (i.e., regulatory settings) along with genetic state (i.e., SignalGP program and environmental-cue-to-tag mapping).
All further work with harvested strains was conducted under environmental conditions identical to that of the treatment they evolved in.

To analyze the relative fitness of knockout strains versus wild type, we seeded 20 $60 \times 60$ toroidal grids with ten cells of each strain, including epigenetic regulator state.
We ran competition experiments for the duration of one snapshot.
Seeded cells generally proliferated to completely fill the toroidal grid in the first quarter of the snapshot.
Competition experiment outcomes were determined by strains' relative cell populations within the grid at the end of the snapshot.

To perform phenotypic comparisons between knockout strains and wild type, we seeded ten cells of each strain onto separate $60 \times 60$ toroidal grids and then cultured them for the duration of one snapshot.

\subsection{Implementation} \label{sup:implementation}

We implemented our experimental system using the Empirical library for scientific software development in C++, available at \url{https://github.com/devosoft/Empirical} \citep{charles_ofria_2019_2575607}.
The code used to perform and analyze our experiments, our figures, data from our experiments, and a live in-browser demo of our system is available via the Open Science Framework at \url{https://osf.io/g58xk/}.
Most replicates finished within a day, but some took up to a week to complete.

\subsection{Reproductive Cooperation} \label{sup:reproductive-cooperation}

\input{fig/reproduction_surrounded.tex}

\subsection{Resource Sharing} \label{sec:resource-sharing}

\input{fig/sharing.tex}
\input{fig/sharing_channelmate.tex}

Figure \ref{fig:sharing} overviews evolved resource sharing behavior across cellular contexts.

Replicates in the flat-wave treatment exhibit an especially elevated rate of resource sharing to cell children.
This could perhaps be due to an especial selective pressure to convey resource towards the group periphery.

Surprisingly, in the Nested-Wave treatment resource was shared at a higher mean rate among high-level same-channel signaling groups than low-level groups.
This observation is likely due to replicates where level-one same-channel signaling groups were composed of single-cell level-zero same-channel signaling groups (where no or very few opportunities for level-zero resource sharing occurred).

Finally, under all treatments resource was transferred to highest-level channelmates at a significantly higher mean rate than to unrelated neighbors (non-overlapping 95\% CI).
This observation suggests that functional cooperation within same-channel groups might have been a common evolutionary outcome under all four treatments.
However, it could potentially be driven exclusively by resource-sharing between direct cellular kin.

Figure \ref{fig:sharing_channelmate} breaks same-channel resource-sharing apart by cellular kin relation.
In all four treatments, mean sharing to direct-kin highest-level channelmates was indeed greater than to other channelmates.
This could be due to an evolutionary incentive to favor direct cell kin over other channelmates, group-level selection for asymmetric resource flow achieved by preferential sharing, or some combination of the two.
However, in all four treatments mean sharing to non-direct-kin highest-level channelmates was also significantly greater than resource sharing to unrelated neighbors (non-overlapping 95\% CI).
Thus, all four treatments appear to be sufficient to select for functional cooperation among cells.

\subsection{Case Study: Morphology} \label{sec:morphology}

\input{fig/ko-morphology.tex}

One of the more striking examples of genetically-encoded same-channel signaling network patterning, in which level-zero same-channel signaling groups arranged as elongated single-file strings, arose in a Nested-Even treatment replicate.
Figure \ref{fig:morphology-wt} provides a snapshot of this strain's same-channel signaling morphology.
Knocking out intracell messaging disrupts the stringy arrangement of same-channel signaling groups, shown in Figure \ref{fig:morphology-ko}.
Figure \ref{fig:morphology-shape} compares the distribution of cells' level-zero same-channel neighbor counts for level-one groups of size nine or greater.
Wild type cells are significantly less likely to have three or four level-zero same-channel neighbors, as we would expect of single-file strings (non-overlapping 95\% CI).
However, we also observed that wild-type level-zero groups had significantly fewer cells
(WT: mean $2.1$, S.D. $1.5$; messaging knockout: mean $4.3$, S.D. 5.1; $p < 0.001$; bootstrap test).
To determine whether morphological patterning besides smaller group size contributed to observed differences in neighbor count, we compared a dimensionless shape factor describing group stringiness (perimeter divided by the square root of area) between the wild type and messaging knockout strains.
Between level-zero group size four (the smallest size stringiness can emerge at on a grid) and level-zero group size nine (the largest size we had replicate wild type observations for), wild type exhibited significantly greater stringiness
(4: $p < 0.01$, bootstrap test; 5: $p < 0.01$, bootstrap test; 6, 7, 8: non-overlapping 95\% CI; 9 $p < 0.01$, bootstrap test).

However, competition experiments between the wild type and knockout strain failed to establish a fitness differential ($6/20$).
Thus, it seems this trait emerged either by drift, as the genetic background of a selective sweep, or --- perhaps less likely --- was advantageous against a divergent competitor earlier in evolutionary history.

\subsection{Case Study: Cell-Cell Messaging} \label{sec:intergroup}

\input{fig/ko-intermessaging-intergroup_border.tex}

Figure \ref{fig:intermessaging-intergroup_border-phen} compares the cell-cell messaging, resource sharing, and parent-propagule phenotypes between wild type and cell-cell messaging knockout variants of a strain evolved under the Nested-Wave treatment.
Cell-cell messaging volume appears generally uniform in the interiors of same-channel signaling groups, but some group-group borders --- largely, but not entirely parent-propagule interfaces --- manifest somewhat depressed cell-cell messaging overlaid with an alternating motif of elevated cell-cell messaging.
We affirmed the adaptiveness of cell-cell messaging in this strain through competition experiments between wild type and knockout variants ($19/20$; $p < 0.001$; two-tailed exact test).
The gene activated by cell-cell messaging in this strain contains a share resource instruction and, indeed, we observed significantly greater net resource sharing in the wild type strain
(%
WT: mean 0.27, S.D. 0.03, $n=20$;
KO: mean 0.23, S.D. 0.02, $n=20$;
$p < 0.001$, bootstrap test
non-overlapping 95\% CI%
). 
However, that same gene also contains a reproduction-inhibiting instruction, leading us to investigate whether cell-cell messaging could influence a broader set of phenotypic traits.

Cell-cell messaging in the wild type strain appears to be associated with a drawn out same-channel signaling network life history.
The wild-type strain exhibits significantly greater mean cell age
(%
WT: mean 59, S.D. 7, $n=20$;
KO: mean 49, S.D. 4, $n=20$;
$p < 0.001$, bootstrap test%
) 
and, across propagule-generation events, significantly greater mean parent group age
(%
WT: mean 1055, S.D. 82, $n=20$;
KO: mean 924, S.D. 62, $n=20$;
$p < 0.001$, bootstrap test%
). 
This strain exhibits the ``sweep'' life history depicted in Figure \ref{fig:lifecycle-sweep}, so propagule generation can be largely or entirely destructive to the parent group.
So, the increase in mean cell age could plausibly be attributable to delayed propagule genesis or, alternatively, delayed propagule genesis could arise from other factors retarding life history.

In this strain, we anecdotally observed that contiguous bands of low cell turnover and anomalous cell-cell messaging volumes frequently arose along parent-propagule borders, but also occasionally between other same-channel signaling network groups.
Cell-cell messaging not only enables functional coordination within cellular collectives but could also enable adaptive communication among cellular collectives.
This possibility motivated us to test for non-uniform interactions between non-parent-propagule same-channel signaling network groups.

We measured mean border age (equivalent to the youngest age of either flanking cell) along the borders of non-parent-propagule same-channel signaling network groups.
Figure \ref{fig:intermessaging-intergroup_border-borderage} splits this statistic out between borders that were disrupted either by cells birthed from members of the same-channel signaling networks flanking the border (``affiliate'') or from a member of a third same-channel signaling network (``neighbor'').
In both wild type and knockout strains, there was significantly more recent turnover in the absence of intrusion by a third same-channel signaling network (non-overlapping 95\% CI, bootstrap test).
Restated, borders invaded by a third party were more on average more stable than those perturbed by either of the flanking same-channel signaling networks.

This phenomenon was accentuated in the wild type strain.
Although the wild type strain exhibits slightly higher turnover rates on borders plied by only two groups, borders invaded by a third group are significantly more stale than the knockout strain (non-overlapping 95\% CI, bootstrap test).

Greater age of borders disrupted by a third party would be consistent with a general slowing of turnover as same-channel signaling networks age or overall reduced resource availability due to the presence of a third party.
However, a primitive tit-for-tat policy where a subset of non-parent-propagule same-channel signaling network borders stabilize (until invaded by a third party) could also contribute to such an observation.

So, does the cell reproduction rate fluctuate uniformly across a same-channel signaling network's borders or can reproduction rate differ significantly between a group's non-parent-propagule neighbor groups at a single time point?
To assess this question, we used Kruskal-Wallis tests (with Bonferroni correction) to screen for same-channel signaling network groups with border reproduction rate distributions that differed significantly between neighboring non-propagule-parent same-channel signaling network groups.
For each same-channel signaling network group, we calculated mean per-border-cell birth rate at the interface of each of its non-propagule-parent neighbor groups.
We collected observations with respect to each neighbor group every eighth update over 256 updates.
Groups with significantly differentiated border reproduction rate distributions occurred in both the wild type and messaging knockout strains.
That is, in both strains, we observed some groups that preferentially expended resource to reproduce at their interfaces with a subset of non-parent-propagule same-channel signaling network group neighbors.

Again, this phenomenon was accentuated in the wild type strain.
A significantly higher proportion of groups exhibited asymmetric border reproduction rates with non-parent-child groups
(%
WT: mean 0.36, S.D. 0.06, $n=20$;
KO: mean 0.28, S.D. 0.04, $n=20$;
$p < 0.001$, bootstrap test%
). 

Messaging between cells registered to different parent-propagule same-channel signaling network groups seems unlikely to directly underlie asymmetric border reproduction rates because execution of the gene targeted by messages triggers resource-sharing to the sender, which we seldom observed between non-parent-propagule groups.
So, intercell messaging within same-channel signaling network groups is most likely responsible.
It seems most plausible that increased incidence of asymmetric border reproduction rates arises as a knock-on effect of the life history retarding demographic effects of cell-cell messaging originally discussed.
Perhaps older, ``full-grown'' same-channel groups arrive at a low-reproduction detente at interfaces with other older, ``full-grown'' same-channel groups while resisting incursion at interfaces with younger, growing same-channel groups.
This would constitute a contextually-expressed tit-for-tat policy, perhaps mediated by cell age or cell generations elapsed from the group's founding propagule cell.

%% file: fig/signalgp-dishtinygp.tex
\begin{figure}
\begin{center}

\hspace*{\fill}%
\begin{minipage}[t]{\columnwidth}
\centering
\vspace{0pt} 
\begin{subfigure}[b]{\textwidth}
\includegraphics[width=\textwidth]{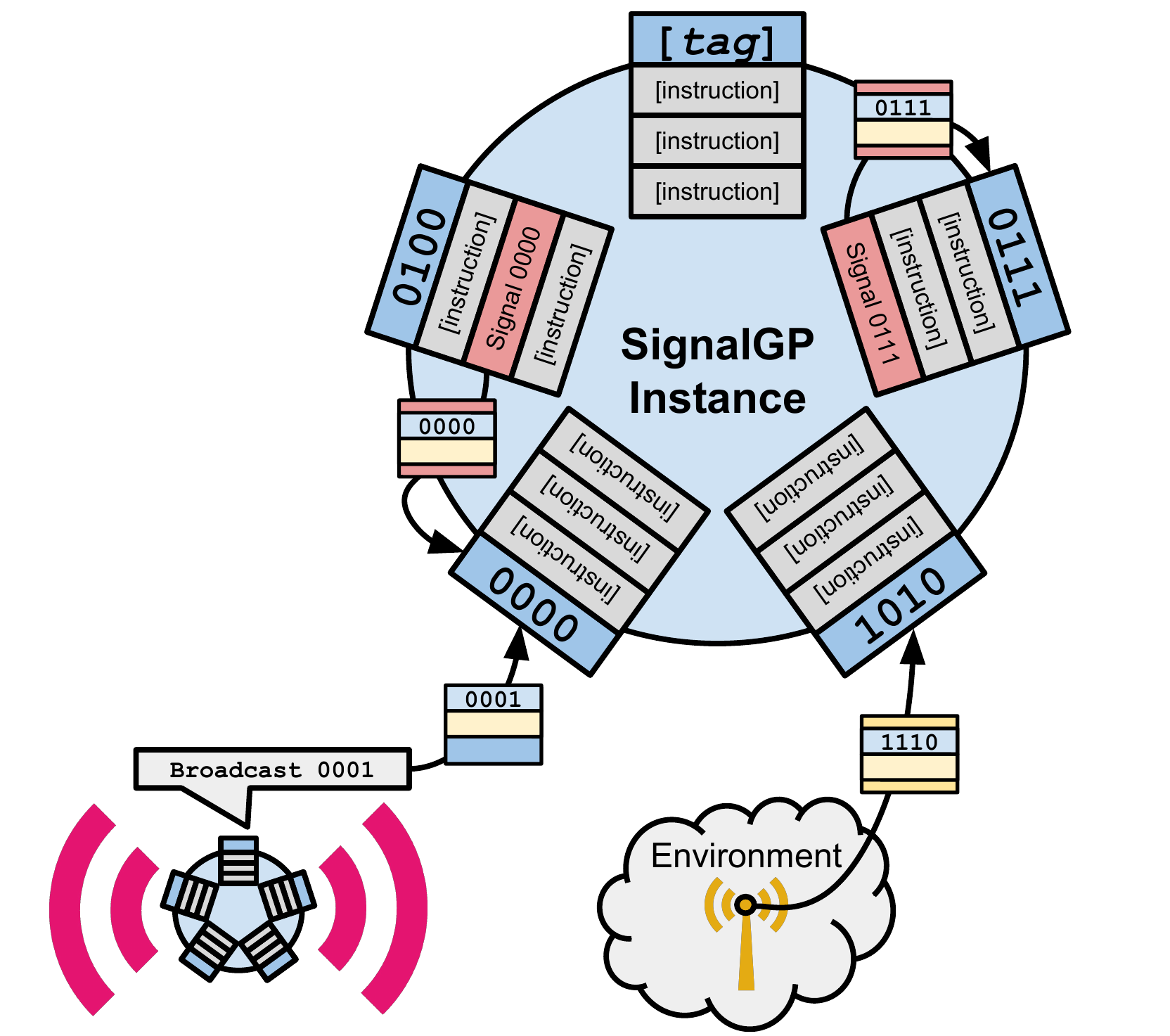}%
\caption{
A cartoon overview of a single SignalGP instance.
SignalGP program modules execute pseudo-concurrently in response to tagged signals, which can originate internally, from the environment, or from other agents.
}
\label{fig:signalgp-cartoon}
\end{subfigure}
\end{minipage}%
\hspace*{\fill}

\hspace*{\fill}%
\begin{minipage}[t]{\columnwidth}
\centering
\vspace{0pt} 
\begin{subfigure}[b]{\textwidth}
\includegraphics[width=\textwidth]{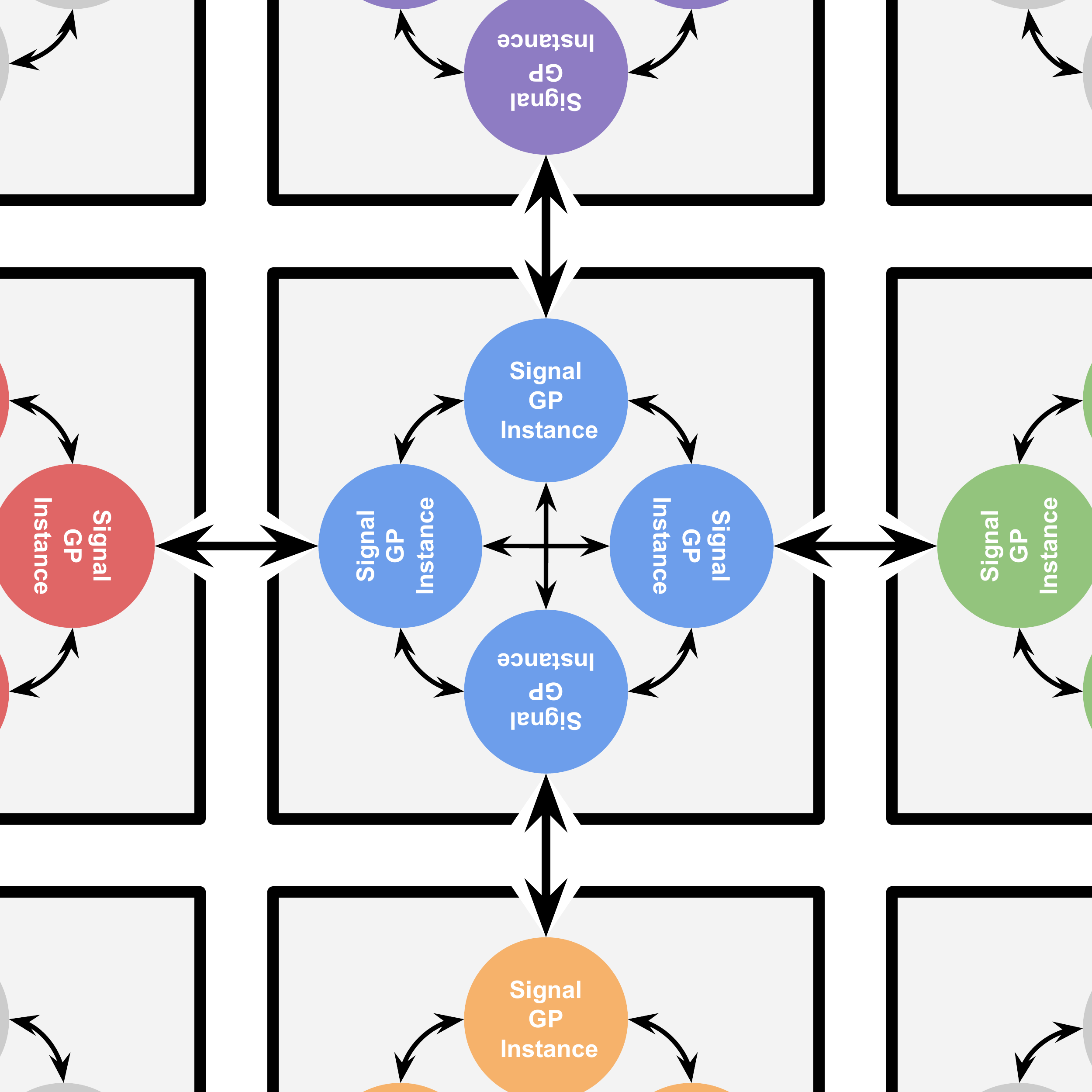}
\caption{
A cartoon overview of how individual SignalGP instances are organized into DISHTINY cells.
Within each DISHTINY cell, each of four independent instances senses environmental state, receives intercellular messages, and determines cell behavior with respect to a single cardinal direction.
All four instances sense non-directional environmental cues and non-directional actions may be taken by any instance.
Instances within a cell communicate via intracellular messaging.
}
\label{fig:dishtinygp-cartoon}
\end{subfigure}
\end{minipage}%
\hspace*{\fill}

\caption{
Schematic illustrations of how individual SignalGP instances function and how individual SignalGP instances are organized into DISHTINY cells.
Figure \ref{fig:signalgp-cartoon} provided courtesy Alexander Lalejini.
}
\label{fig:signalgp-dishtinygp}
\end{center}
\end{figure}

%% file: fig/productivity.tex
\begin{table*}[!htbp]
\begin{center}

\begin{tabular}{l|c|c|c|c}%
\bfseries Measure
  & \bfseries Flat-Even
  & \bfseries Flat-Wave
  & \bfseries Nested-Even
  & \bfseries Nested-Wave
\csvreader[head to column names]{productivity.csv}{}
{\\\hline\Measure
  & \FlatEven
  & \FlatWave
  & \NestedEven
  & \NestedWave
}
\end{tabular}

\caption{
Observed productivity at epoch 1 (mean $\pm$ S.D.)
}
\label{tab:productivity}
\end{center}
\end{table*}

%% file: fig/systematics.tex
\begin{sidewaystable}[!htb]
\begin{center}

\begin{tabular}{l|c|c|c|c|c|c|c|c}%
&\multicolumn{4}{c|}{Epoch 1}
&\multicolumn{4}{c}{Epoch 4}\\
\bfseries Measure
  & \bfseries Flat-Even
  & \bfseries Flat-Wave
  & \bfseries Nested-Even
  & \bfseries Nested-Wave
  & \bfseries Flat-Even
  & \bfseries Flat-Wave
  & \bfseries Nested-Even
  & \bfseries Nested-Wave
\csvreader[head to column names]{systematics.csv}{}
{\\\hline\Measure
  & \FlatEvenShort
  & \FlatWaveShort
  & \NestedEvenShort
  & \NestedWaveShort
  & \FlatEvenLong
  & \FlatWaveLong
  & \NestedEvenLong
  & \NestedWaveLong
}
\end{tabular}

\caption{
Systematics outcomes (mean $\pm$ S.D.)
}
\label{tab:systematics}
\end{center}
\end{sidewaystable}

%% file: fig/reproduction_surrounded.tex
\begin{figure}[!htbp]
\begin{center}

\includegraphics[width=\columnwidth]{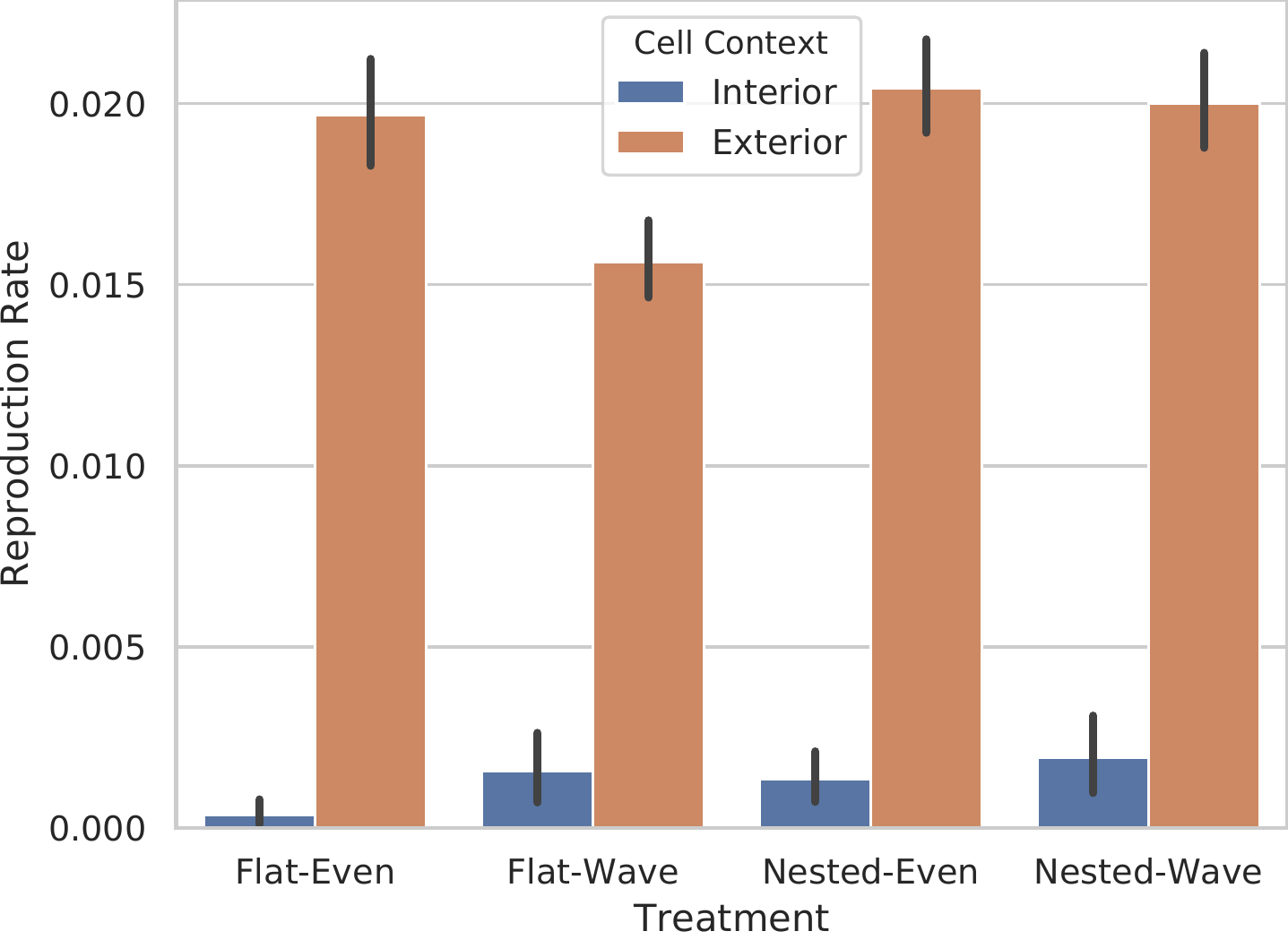}

\caption{
Cellular reproduction rates at the interior or exterior of highest-level same-channel signaling networks.
Error bars indicate 95\% confidence.
}
\label{fig:reproduction_surrounded}
\end{center}
\end{figure}

%% file: fig/sharing.tex
\begin{figure*}[!htbp]
\begin{center}

\includegraphics[width=\textwidth]{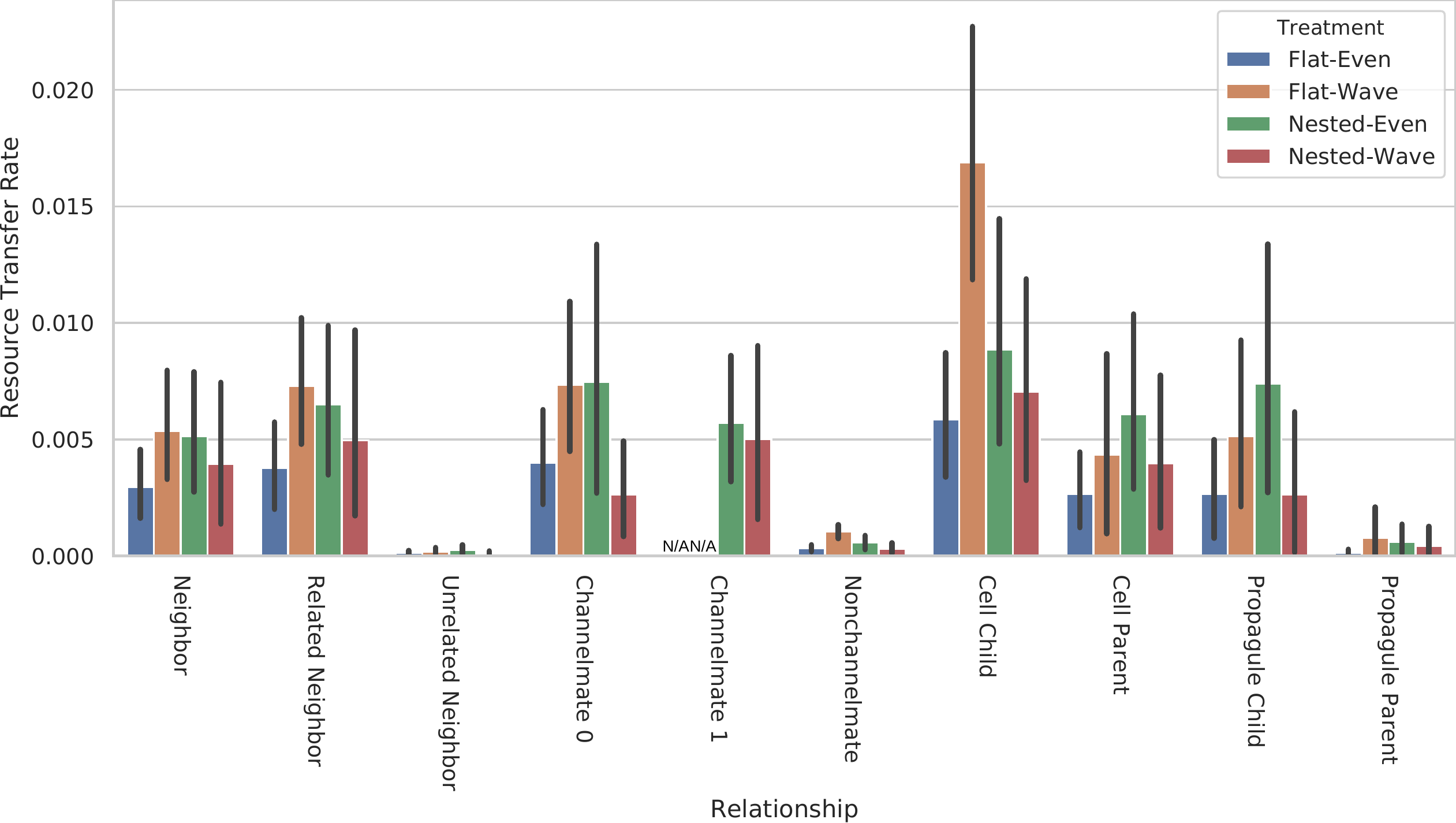}

\caption{
Resource sharing rates across donor-recipient relationships.
Neighbor describes any potential recipient cell.
Related neighbor describes a recipient cell that is a direct cellular progenitor or offspring of the donor, registered to a same signaling channel as the donor, or a member of a signaling channel that is a progenitor or offspring of the donor's.
Unrelated neighbors constitutes all neighbors that are not related neighbors.
Channelmate refers to donor-recipient pairs that are registered to a same signaling channel.
Note that level-one groups are not defined in the flat treatment.
Non-channelmate recipients are not registered to any same signaling channel in common with the potential donor.
Cell child and parent describe direct nuclear cell relationships between donor and recipient.
Finally, a propagule child relationship exists when a donor cell is a member of the highest-level signaling channel that directly begat the recipient's.
A propagule parent relationship describes the converse.
Error bars indicate 95\% confidence.
}
\label{fig:sharing}
\end{center}
\end{figure*}

%% file: fig/sharing_channelmate.tex
\begin{figure*}[!htbp]
\begin{center}

\includegraphics[width=\textwidth]{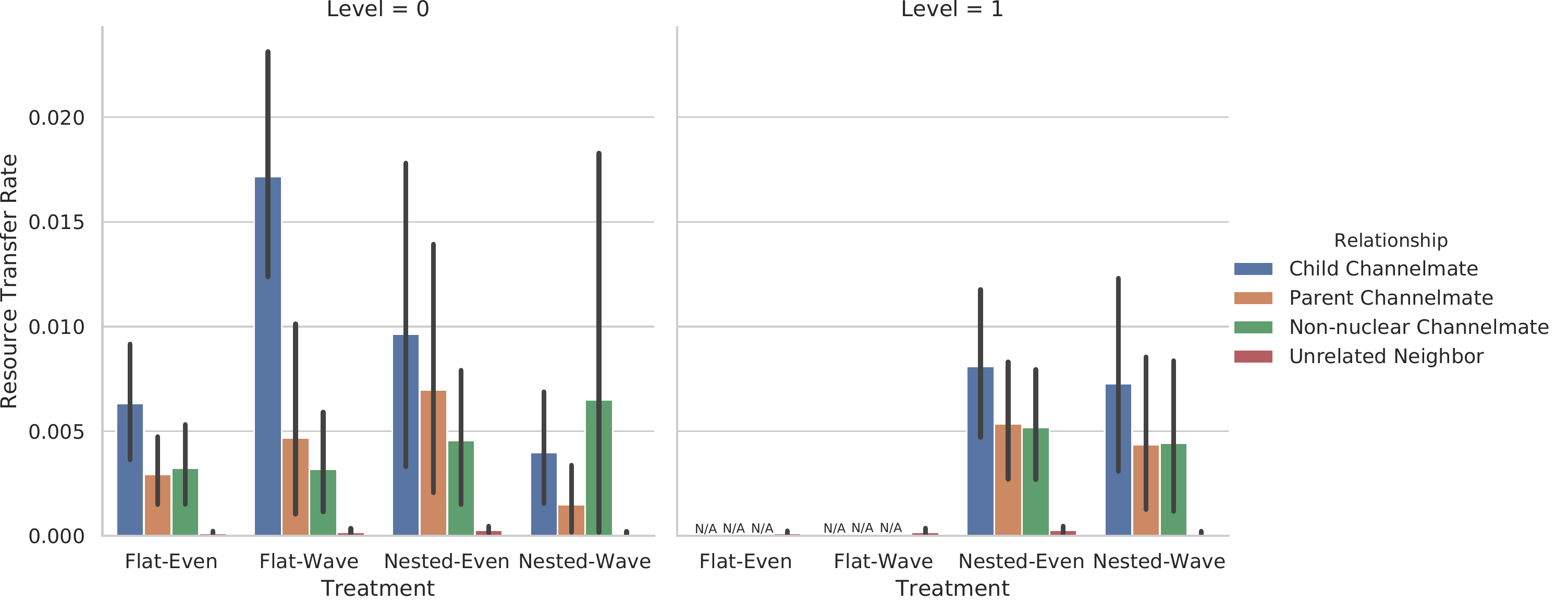}

\caption{
Resource sharing to mutually exclusive sub-categories same-channel cellular neighbors: cellular child, cellular parent, and neither (``non-nuclear'').
Resource sharing to entirely non-related cells (no cell, channel, or propagule relation) is included for comparison.
Note that level-one groups are not defined in either of the flat treatments.
Error bars indicate 95\% confidence.
}
\label{fig:sharing_channelmate}
\end{center}
\end{figure*}

%% file: fig/ko-morphology.tex
\begin{figure}[!htbp]
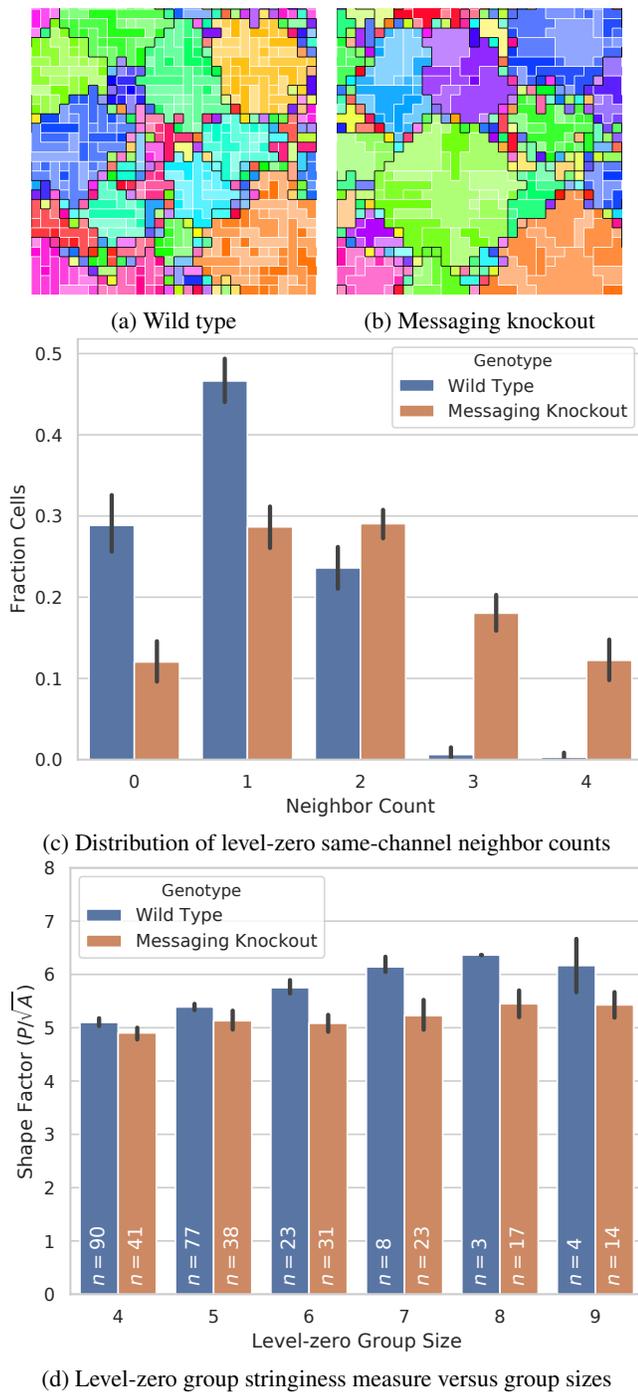

\begin{center}

\hspace*{\fill}%
\begin{minipage}[t]{0.45\columnwidth}
\centering
\vspace{0pt} 
\begin{subfigure}[b]{\textwidth}
\adjincludegraphics[width=\textwidth, trim={{.0\width} {.0\width} {.5\width} {.5\width}}, clip]{knockout/morphology/wildtype/seed=1+title=channel_viz+treat=resource-even__channelsense-yes__nlev-two+update=8188+_data_hathash_hash=cb64cdf045bc6049+_script_fullcat_hash=7e789c981e3d0e4f+_source_hash=53a2252-clean+ext=}
\caption{Wild type}
\label{fig:morphology-wt}
\end{subfigure}
\end{minipage}%
\hfill
\begin{minipage}[t]{0.45\columnwidth}
\centering
\vspace{0pt} 
\begin{subfigure}[b]{\textwidth}
\adjincludegraphics[width=\textwidth, trim={{.0\width} {.0\width} {.5\width} {.5\width}}, clip]{knockout/morphology/knockout/seed=1+title=channel_viz+treat=resource-even__channelsense-yes__nlev-two+update=8188+_data_hathash_hash=9a4119947348e91d+_script_fullcat_hash=7e789c981e3d0e4f+_source_hash=53a2252-clean+ext=}%
\caption{Messaging knockout}
\label{fig:morphology-ko}
\end{subfigure}
\end{minipage}%
\hspace*{\fill}

\hspace*{\fill}%
\begin{minipage}[t]{\columnwidth}
\centering
\vspace{0pt} 
\begin{subfigure}[b]{\textwidth}
\adjincludegraphics[width=\textwidth]{knockout/morphology/title=group_shape+_data_hathash_hash=cb1733796dea778f+_script_fullcat_hash=68cf35a1759c64ac+_source_hash=53a2252-clean+ext=}
\caption{Distribution of level-zero same-channel neighbor counts}
\label{fig:morphology-shape}
\end{subfigure}
\end{minipage}%
\hspace*{\fill}

\hspace*{\fill}%
\begin{minipage}[t]{\columnwidth}
\centering
\vspace{0pt} 
\begin{subfigure}[b]{\textwidth}
\adjincludegraphics[width=\textwidth]{knockout/morphology/title=group_perimeter_area+_data_hathash_hash=b02d4442d68976b7+_script_fullcat_hash=4198d7d7c0b9f172+_source_hash=53a2252-clean+ext=}
\caption{Level-zero group stringiness measure versus group sizes}
\label{fig:morphology-factor}
\end{subfigure}
\end{minipage}%
\hspace*{\fill}

\caption{
Comparison of a wild type strain evolved under the ``Nested-Even'' treatment with stringy level-zero same-channel signaling networks and the corresponding intracellular-messaging knockout strain.
Subfigures \ref{fig:morphology-wt} and \ref{fig:morphology-ko} visualize same-channel signaling network layouts;
color hue denotes and black borders divide level-one same-channel signaling networks while
color saturation denotes and white borders divide level-zero same-channel signaling networks.
Subfigures \ref{fig:morphology-shape} and \ref{fig:morphology-factor} quantify the morphological effect of the intracellular-messaging knockout.
Error bars indicate 95\% confidence.
View an animation of the wild type strain at \url{https://mmore500.com/hopto/q}.
View the wild type strain in a live in-browser simulation at \url{https://mmore500.com/hopto/f}.
}
\label{fig:ko-morphology}
\end{center}
\end{figure}

%% file: fig/ko-intermessaging-intergroup_border.tex
\begin{figure}[!htbp]
\begin{center}

\centering
\hspace*{\fill}%
\begin{minipage}[t]{0.05\columnwidth}
\vspace{0pt} 
\rotatebox{90}{Messaging}%
\end{minipage}%
\hfill
\begin{minipage}[t]{0.45\columnwidth}
\centering
\vspace{0pt} 
\adjincludegraphics[width=\textwidth, trim={{.66\width} {.66\width} {.0\width} {.0\width}}, clip]{knockout/intermessaging-intergroup_border/wildtype/seed=1+title=directional_messaging_viz+treat=resource-wave__channelsense-yes__nlev-two+update=7168+_data_hathash_hash=3895dfa0dd602b4c+_script_fullcat_hash=6b7e0389992dd616+_source_hash=53a2252-clean+ext=}%
\end{minipage}%
\hfill
\begin{minipage}[t]{0.45\columnwidth}
\centering
\vspace{0pt} 
\adjincludegraphics[width=\textwidth, trim={{.66\width} {.66\width} {.0\width} {.0\width}}, clip]{knockout/intermessaging-intergroup_border/knockout/seed=1+title=directional_messaging_viz+treat=resource-wave__channelsense-yes__nlev-two+update=7168+_data_hathash_hash=24546cc614406803+_script_fullcat_hash=6b7e0389992dd616+_source_hash=53a2252-clean+ext=}%
\end{minipage}%
\hspace*{\fill}


\hspace*{\fill}%
\begin{minipage}[t]{0.05\columnwidth}
\vspace{0pt} 
\rotatebox{90}{Resource Sharing}%
\end{minipage}%
\hfill
\begin{minipage}[t]{0.45\columnwidth}
\centering
\vspace{0pt} 
\adjincludegraphics[width=\textwidth, trim={{.66\width} {.66\width} {.0\width} {.0\width}}, clip]{knockout/intermessaging-intergroup_border/wildtype/seed=1+title=directional_sharing_viz+treat=resource-wave__channelsense-yes__nlev-two+update=7172+_data_hathash_hash=3895dfa0dd602b4c+_script_fullcat_hash=3a1e851383e0ffd4+_source_hash=53a2252-clean+ext=}%
\end{minipage}%
\hfill
\begin{minipage}[t]{0.45\columnwidth}
\centering
\vspace{0pt} 
\adjincludegraphics[width=\textwidth, trim={{.66\width} {.66\width} {.0\width} {.0\width}}, clip]{knockout/intermessaging-intergroup_border/knockout/seed=1+title=directional_sharing_viz+treat=resource-wave__channelsense-yes__nlev-two+update=7172+_data_hathash_hash=24546cc614406803+_script_fullcat_hash=3a1e851383e0ffd4+_source_hash=53a2252-clean+ext=}%
\end{minipage}%
\hspace*{\fill}


\vspace{1.0ex}

\hspace*{\fill}%
\begin{minipage}[t]{0.05\columnwidth}
\vspace{0pt} 
\end{minipage}%
\hfill
\begin{minipage}[t]{0.45\columnwidth}
\centering
\vspace{0pt} 
Wild Type
\end{minipage}%
\hfill
\begin{minipage}[t]{0.45\columnwidth}
\centering
\vspace{0pt} 
Messaging Knockout
\end{minipage}%
\hspace*{\fill}

\vspace{1.0ex}

\begin{subfigure}{\columnwidth}
  \caption{Phenotype visualizations}
  \label{fig:intermessaging-intergroup_border-phen}
\end{subfigure}

%
%




\hspace*{\fill}%
\begin{minipage}[t]{0.8\columnwidth}
\centering
\vspace{0pt} 
\begin{subfigure}[b]{\textwidth}
\includegraphics[width=\textwidth]{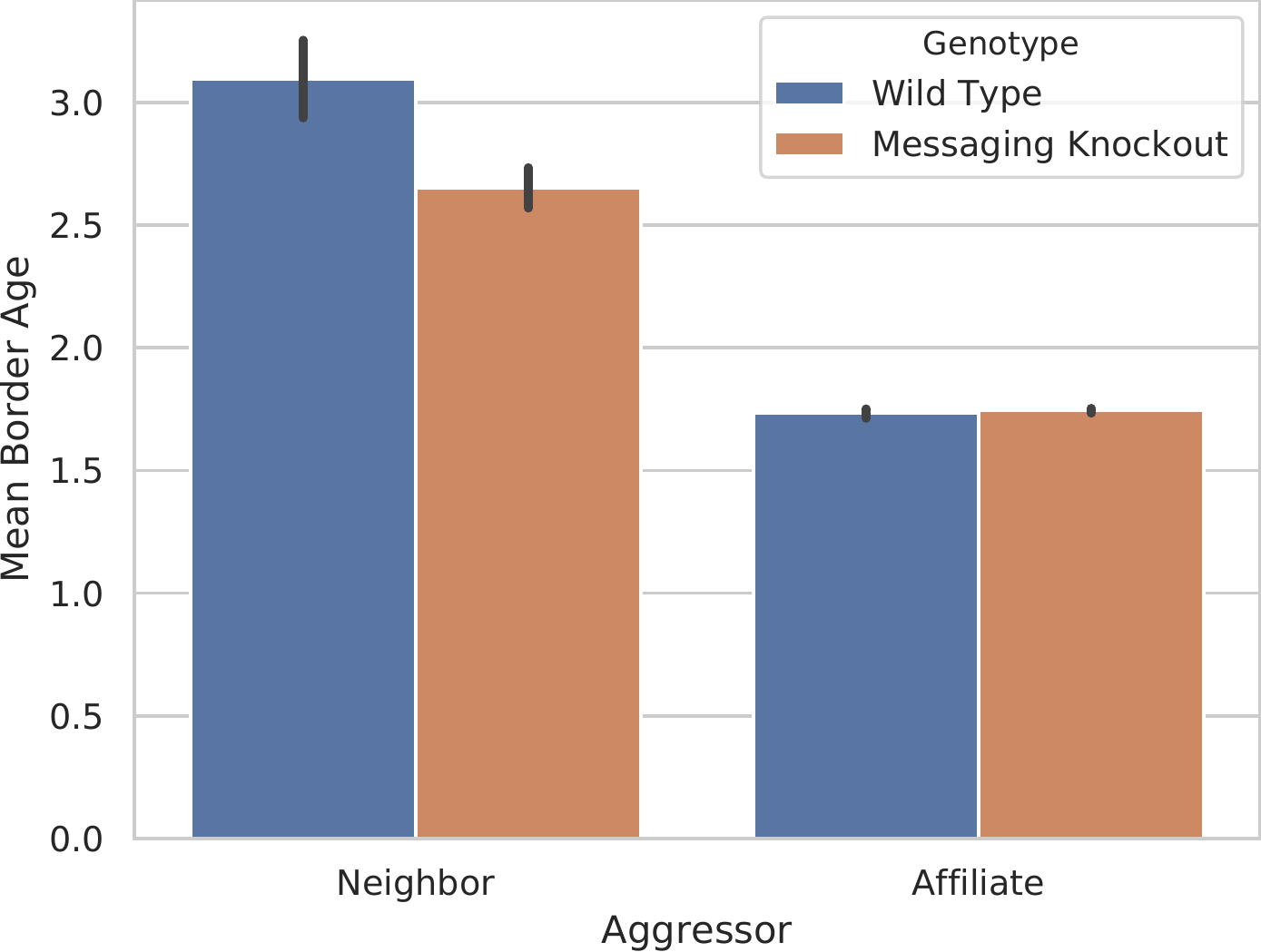}%
\caption{Border age}
\label{fig:intermessaging-intergroup_border-borderage}
\end{subfigure}
\end{minipage}%
\hspace*{\fill}

\vspace{1ex}


\caption{
Comparison of wild type strain evolved under the ``Nested-Wave'' treatment and corresponding intercell messaging knockout strain.
Subfigure \ref{fig:intermessaging-intergroup_border-phen} visualizes phenotypic traits in the wild type and knockout strain.
In the messaging visualization, color coding represents the volume of incoming messages.
White represents no incoming messages and the magenta to blue gradient runs from one incoming message to the maximum observed incoming message traffic.
In the resource sharing visualization, this same color coding represents the amount of incoming shared resource.
Solid black borders divide level-one same-channel signaling networks and dotted light gray borders divide level-zero same-channel signaling networks.
Subfigure \ref{fig:intermessaging-intergroup_border-borderage} quantifies knockout effects on border age.
View an animation of the wild type strain at \url{https://mmore500.com/hopto/o}.
View the wild type strain in a live in-browser simulation at \url{https://mmore500.com/hopto/d}.
Error bars indicate 95\% confidence.
}
\label{fig:ko-intermessaging-intergroup_border}

\end{center}
\end{figure}